**Cover Page**

On the origin of $^7$Be isotopic records in a Calcium, Aluminium, -rich inclusion

Ritesh Kumar Mishra† and Kuljeet Kaur Marhas

Present address: Klaus-Tschira-Labor für Kosmochemie, Institut für

Geowissenschaften, Im Neuenheimer Feld 234-236 Ruprecht-Karls-Universität,

Heidelberg, D-69210 Germany.

Ritesh Kumar Mishra

Affiliations:

Center for Isotope Cosmochemistry and Geochronology, Astromaterials Research and

Exploration Science division, EISD-XI, NASA-Johnson Space Center, 2101, NASA

Parkway, Houston, TX 77058, USA.

Ritesh Kumar Mishra

Planetary Sciences Division, Physical Research Laboratory, Navrangpura,

Ahmedabad, Gujarat, 360009 India.

Kuljeet Kaur Marhas

Format Letter

Word counts:

    Introductory paragraph: 241

    Main Text: 2174

    Methods: Nil

    Legends: Fig 1, Fig 2

Number of references: 32 +(14 Supplementary)

Number of Figures: 2

Number of Tables: 1

Supplementary information : words 5321 Figure: S1-6  Table  S1

Abstract:

**A prime question in the formation and early evolution of the Solar system studies is to discern the source(s) of short-lived now extinct nuclides and to determine the ab-initio isotopic composition of our Solar System[1]. The proposed genesis of a short-lived now extinct radionuclide, $^{10}$Be, by spallation reactions of carbon and oxygen led to the hypothesis of enhanced irradiation in the early Solar system[2-8]. An alternative scenario of production of $^{10}$Be ($t_{1/2}$ =1.386 ± 0.016 million years[9]) by "$\nu$" process in a low mass star (11.8$M_\odot$) core collapse supernova has been recently suggested[10] that can explain the observed abundance of $^{10}$Be in the early Solar System. Here, we report well resolved excesses in $^{7}$Li/$^{6}$Li of up to ~21.5 % in a Type B1 Ca,-Al rich inclusion (CAI) from the Efremovka meteorite that correlate with $^{9}$Be/$^{6}$Li, suggestive of *in situ* decay of $^{7}$Be. The *in situ* decay of $^{7}$Be, with characteristic half-life of 53.12± 0.07 days[11] to $^{7}$Li, entails multiple episodes of enhanced irradiation in the ESS. The short half-life of $^{7}$Be limits its production by interaction of Solar energetic particles with the nebular gas and solids and provides constraints on genealogy and chronology of CAIs. Irradiation of precursor solids/gas of CAIs of Solar composition by a superflare ($L_x \approx 10^{32}$ erg/sec) during the terminal phase of class I or II of pre-main sequence stages of the Sun cogently explains the isotopic properties, distinctive petrographic features, and diffusivity constraints in the CAI.**

Sun-like stars during their pre-main sequence stages emanate orders of magnitude higher X-, EUV emissions[12,13]. These highly energetic, high temperature events operating on large spatial dimensions (up to a few astronomical units) occur over variable time scales (few days to hundreds of ka) and have causal linkage to collapse of molecular clouds and formation of proto-planetary disks and planetary systems. Theoretical models and astronomical observations have furthered our understanding of the stochastic processes and dependence on the *ab initio* conditions on the formation of the planetary systems. However, these studies cannot constrain to the unique scenario of events during the formation of our Solar System. Fossil records of short-lived now-extinct radionuclides (SLNs) in meteorites help glean this crucial high resolution temporal information. One of the major posited sources of the SLNs is spallation reactions triggered by the energetic events. SLNs with half-lives ≤10 million years ($10^6$ Yrs =Ma=Myr) therefore provide crucial specific information about the earliest stage events, processes, and astrophysical environments in the early solar system. Until recently, $^{10}$Be has been advocated as the quintessential radionuclide sourced mostly from irradiation/spallation of carbon and oxygen nuclides either in the early Solar protoplanetary disk or in the inter-stellar medium[2-8,14]. The assertion rested squarely on the following two facts: (i) the short half-life of $^{10}$Be of 1.386± 0.016 million years[9] (Ma) mandated a maximum time interval of ≤7 Ma between production and isotope systematics closure; and (ii) low binding energy (~6 MeV/nucleon) of beryllium nuclides resulted in it being consumed as fuel in any high temperature stellar environment. In contravention to this view, it has been recently argued and demonstrated that in a low mass (≤11.8 $M_\odot$) core collapse supernova (CCSN), the neutrino spallation $^{12}$C(ν,ν'pp)$^{10}$Be reaction produces sufficient $^{10}$Be to mostly explain observed abundances in CAIs[10]. Their model calculations can simultaneously also account for canonical abundance of $^{41}$Ca, $^{107}$Pd, and $^{60}$Fe in the early solar system. These model calculations therefore significantly weaken the tenet of production of $^{10}$Be by irradiation 'only' and by inference evidence of irradiation in the ESS[2-8]. A plausible presence for $^7$Be in only 'one' Allende type B1 CAI (3529-41) has been suggested[2]. A more rigorous method of calculating the cosmogenic corrections for the obtained isotopic data in Allende CAI (3529-41) showed significantly lower excesses in a few fassaite (up to ~31.6 ‰) and anorthite (up to ~43‰) while the cosmogenically corrected isotopic ratios remained the same in the melilite. Despite these large corrections, the initial $^7$Be/$^9$Be value inferred by regression of data remained compatible, within uncertainties, with the initial studies that had obtained an isochron with $^7$Be/$^9$Be slope of (6.1±1.3) ×10$^{-3}$ (2σ) with non-chondritic initial $^7$Li/$^6$Li$_0$ of 11.49±0.13 (2σ). Crucially though, statistical evaluation of distribution of data and residues suggested that the correlation between $^7$Li/$^6$Li and $^9$Be/$^6$Li is non linear and/or the data has variations beyond the associated uncertainities[15].

Refractory mineral phases predicted to form in a cooling gas of solar composition are primarily oxides and silicates of calcium, aluminium, titanium, and magnesium and are found in objects referred to as calcium-, aluminium rich inclusions (CAIs)[16]. The absolute U-Pb chronometer dates formation of CAIs within a few million years at ~4568 Ma[17] which is taken as the "time zero" of the formation of the solar system. Carbonaceous chondrites of Vigarano type (CV), have abundant (~8 volume%) CAIs and show a metamorphic sequence[16]. Efremovka (CV reduced type), classified as petrographic type ~3.1-3.4, is one of the least altered meteorites[18] where primary/pristine features and evidence of nebular and parent body metamorphism have been well documented amongst its different components.

Lithium, beryllium, and boron *in situ* isotopic studies were carried out in melilite, in a CAI from the Efremovka meteorite using secondary ion mass spectrometry[4] (see Methods). E40 is a type B1 CAI of ~6.4×4.8 mm whose petrography, mineralogy, and $^{26}$Al-$^{26}$Mg isotope systematics has been reported previously[19] (Supplementary Figure 1 A-C, Methods). Lithium and beryllium concentrations in the melilite in CAI range from 1-262 ppb, and 39-607 ppb, respectively (Methods, additional details in supplementary information). Lithium *in situ* isotopic studies carried out in the CAI from Efremovka show range of excess in $^7$Li/$^6$Li up to 215±110 ‰ (2σ) in E40 (Table 1) after correction for cosmogenic contributions. $^7$Li/$^6$Li ratios after cosmogenic corrections that linearly correlate with $^9$Be/$^6$Li are suggestive of *in situ* decay of $^7$Be. The internal isochron of E40 obtained by error weighted regression (model 1 of Isoplot 4) of data from melilite yields $^7$Be/$^9$Be ratio of (1.2±1.0)×10$^{-3}$ (95% conf.) (Fig. 1a). The isochron intercept δ$^7$Li$_0$ of isochron is 0.0±13.3 ‰; consistent within errors with the bulk terrestrial lithium isotopic composition. The resolved isochron is anchored on the strength of two analyses that are completely resolved (>3σ) and weighted mean of the three analyses at the highest $^9$Be/$^6$Li. It is worth mentioning that all analysed regions with high [Be] show higher mean $^7$Li/$^6$Li but position (and probably deviation from expected) on the isochron and error is also determined by a rather wide range of high-low [Li] incorporated. Typically large errors are associated with each measurement because of trace abundance of both lithium and beryllium and therefore while the mean values lie close to the expected trend, only a few are completely resolved from the chondritic (12.02). If the initial $^7$Li/$^6$Li$_0$ were to be ~12.5, instead of 12.02, the data would not support evidence of presence of $^7$Be. The most precise analyses (4/5) with [Li] of greater than 40 ppb however are discordant with the elevated ratio and negate the assumption of existence of elevated bulk lithium isotopic composition of ~12.5. The boron concentration in melilite in E40 shows a range of 3.7-632.0 ppb and yields $^9$Be/$^{11}$B maximum range of up to 35. E40 gives an isochron with a positive slope corresponding to a $^{10}$Be/$^9$Be ratio of (1.6±1.0)×10$^{-3}$ (2σ) (Fig. 1b) and δ$^{10}$B$_0$ of -56.8±60.5 ‰ (2σ). A range of $^{10}$Be/$^9$Be from (3-100)×$10^{-4}$ has been previously observed in CAIs[8]. However, a notable difference is that the observed $^{10}Be/^9Be$ in E40 is ~2 times higher than those observed previously in other CV CAIs[8], typically in the range of (4-8)×$10^{-4}$. The previously measured $^{26}Al/^{27}Al$ ratio interpreted in terms of time in a homogenous ESS with canonical $^{26}Al/^{27}Al$ of (5.25±0.13)×$10^{-5}$ implies that the CAI (E40) with $^{26}Al/^{27}Al$ ratio of (3.4±1.0)×$10^{-5}$ formed at 0.45±0.3Ma. At odds and interestingly, Allende CAI 3529-41 has higher (similar within errors!) $^{26}Al/^{27}Al$ of (4.1± 1.2)×$10^{-5}$, higher $^7Be/^9Be$ of (6.1± 1.3)×$10^{-3}$ but smaller $^{10}Be/^9Be$ of (8.8± 0.6)×$10^{-4}$ (all 2σ) abundances. This dissonance can be explained in two scenarios (1) If the different mean $^{26}Al/^{27}Al$ has chronological meaning, then the two CAIs experienced distinct irradiations events in time. The one occurring earlier was seen by Allende CAI for shorter duration (considering intensity of irradiation constant) or alternatively the irradiation event was lesser in intensity (considering time duration constant). (2) In the other scenario where $^{26}Al$ abundance has no chronological meaning, Allende CAI 3529-41 was produced in proximally to the irradiation source before being entrained in outward transporting wind in short time period.

$^7Be$ decays via electron capture to $^7Li$ with a half-life of 53.12±0.07 days[11]. Such a short half-life precludes its provenance from (1) core collapse supernova (CCSN)[10], and (2) magnetic trapping of either galactic $^7Be$, or galactic cosmic ray produced components by the Solar nebula[14]. Hence, this limits production of $^7Be$ by interaction of solar energetic particles (SEPs) with precursor gas/solids of CAIs, or CAIs itself. The concomitant presence of $^{10}Be$, and $^{26}Al$ in E40 places stringent constraints on (1) contribution of local irradiation towards the inventory of SLNs, (2) duration and/or intensity of the irradiation event, and (3) the time scale of the thermal event leading to formation of this specific CAI. Model calculations of interaction of SEPs with

precursors solids, gas and CAIs under different plausible scenarios (time period, intensity, rigidity of flux, composition of solids) were carried out to match the observed isotopic records and delineate the astrophysical environment[4,20] (Methods). A particular scenario model calculations that closely matches the observed isotopic records, shown in Fig. 2, allow the following inferences: (1) Production of $^7$Be is linearly associated with production of $^{10}$Be (2) Occurrence of a superflare ($L_x \approx 10^{32}$erg/sec) to provide required flux (~$10^{10}$ protons cm$^{-2}$sec$^{-1}$). The correlated production of $^7$Be, and $^{10}$Be is primarily due to the same target (oxygen, carbon nuclie), and similar energy range of maximum reaction cross section. The co-production of $^{7,10}$Be argues in favor of irradiation as the major source of $^{10}$Be. However, since the studied CAIs formed later (~0.45 Ma), and model calculations of production of SLNs are anchored to $^7$Be the study does not specifically negate some minor contribution to $^{10}$Be from CCSN, or an earlier irradiation event in the canonical, or Fractionation and unidentified nuclear effects (FUN) CAIs. Authors note that the FUN CAIs AxCAI 2771 and KT-1 have about an order of magnitude lower abundance of $^{10}$Be/$^9$Be compared to E40 and KT-1 shows a scattered correlation between $^7$Li/$^6$Li and 1/[Li]. The best correspondence between our model calculations and observed isotopic records in CAIs is obtained by considering irradiation of precursors solids with CI (carbonaceous Ivuna; ≈Solar) composition. In this specific model calculation inferred 'canonical' abundance of $^{41}$Ca/$^{40}$Ca and $^{53}$Mn/$^{55}$Mn (within a factor of ~10) is produced while $^{36}$Cl/$^{35}$Cl is under produced by two orders of magnitudes (Fig. 2) within ~(3±2) years. $^{41}$Ca, $^{53}$Mn, $^{36}$Cl, $^{26}$Al, etc. are easily produced by spallation reaction and have been prime candidates for SLNs produced by irradiation. Currently, various lines of evidences argue for $^{26}$Al being produced by stellar nucleosynthesis while $^{36}$Cl being spallogenic. A recent downward revision of abundance of $^{41}$Ca is being suggested to be of more likely spallogenic origin over the previously argued stellar sourced. So the model calculations are in general agreement but for $^{36}$Cl. Isotopic data and petrographic evidences suggest that $^{36}$Cl is produced in the later stage (>1.5 Ma) in a volatile rich environment and is therefore not a major short coming of the calculation but an interesting avenue to further explore and constrain. Production of SLNs in a few irradiation scenarios has been attempted previously that primarily consider abundance of $^{10}$Be/$^{9}$Be of (4-8)×10$^{-4}$ in CV CAIs as the fiducial[21-25]. Superflares with inferred energy have been recently observed for solar type stars[12,13].

The petrography and mineralogy of the CAI (Supplementary Figure 1) provides additional useful information about the prevalent condition, i.e. duration and nature of the thermal event. The melilite rich mantle of E40, and the mineralogical composition (euhedral spinels, rarity of anorthite) suggests it experienced a peak temperature of ~1420°C, and cooled at ~0.5°C/hr[26] (Methods, Supplementary information). Diffusion of lithium within melilite has not been experimentally determined. Diffusivity of lithium, and boron in pyroxene[27], and basaltic melts are amongst the fastest, and two to five orders of magnitude faster than that of magnesium, oxygen, and silicon. Notwithstanding this fact, preservation of isotopic records of faster diffusing species of boron ($^{10}$Be-$^{10}$B systematics) and lithium ($^{7}$Be-$^{7}$Li systematics) are frequently accompanied by disturbed isotopic records of slower diffusing species of magnesium ($^{26}$Al-$^{26}$Mg systematics) in CAIs presenting a conundrum from diffusivity considerations. The limiting constraints from diffusion of lithium, boron are crucial to understand, and model the spatio-temporal evolution, and transport of solids in the proto-planetary disks and thereby illustrate causal linkage with the irradiation events.

A late (~0.5 Ma) superflare with X-ray Luminosity ($L_x$) of $\approx 10^{32}$ erg/sec that irradiates precursor solids of CAI having CI composition can simultaneously explain the isotopic, petrographic, diffusivity constraint, and model calculations. Such a superflare can provide required flux of ~$10^{10}$ proton/cm$^2$/sec near reconnection region/'X' point to produce the observed isotopic abundance in the CAI while creating cosmochemical environment that satisfy other petrographic, diffusivity properties. The short duration intense irradiation of CI composition prescursors is required to simultaneously satisfy the isotopic, petrographic, and diffusion constraints. The $^{26}$Al isochron of E40 requires the occurrence of such a flare at the end of class I, or during class II stage of pre-main sequence evolution. Precision of the bulk (combined) $^{26}$Al isochron of CAIs, and AOAs presently constrain formation of CAIs over a period of ~4000 years[28], and major episode of irradiation during the pre T-Tauri, or T-Tauri phase (Class 0-III; total time >2Ma)[5,7,8,25]. High precision *in situ* internal $^{26}$Al isochrons in CAIs suggest multiple epochs of CAIs' formation including a later one at ~0.5 Ma[29] which supports the observed isotopic records. The late epoch of intense irradiation inferred in the present study is in general agreement with the observed isotopic records of $^{36}$Cl in CAIs and chondrules[30,31]. The higher abundance of $^{26}$Al/$^{27}$Al, and lower abundance of $^{10}$Be/$^9$Be in CV CAIs[8] compared to the E40 CAI studied in the present study taken together with isotopic records of $^7$Be/$^9$Be implies a later (re)occurrence of an event of intense irradiation seen by the CAIs in the present study. Alternatively, precursor solids were swept up in closer proximity to the Sun during this particular later epoch before being ejected outwards. The short period of irradiation during a time period of about a year significantly delimits the constraints from a previous independent study of $^{10}$Be, and $^{51}$V in CAIs which suggested irradiation duration of less than 300 years at 0.1 au[25].

Acknowledgements

Secondary ion mass spectrometer (SIMS) at Physical Research Laboratory, Ahmedabad, India is supported partially by financial grants from Department of Space, Govt. of India. KKM gratefully acknowledges support from Scientific and Engineering Research Board, India (SERB-WES grant# SB/WEA-007/2013) for carrying out the research. The financial support during the preliminary preparation of the manuscript at NASA Johnson Space Center, Houston under NASA post-doctoral program (NPP) fellowship administered by the Oak Ridge National Laboratory associated universities is gratefully acknowledged. Financial support from the Alexander Von Humboldt foundation during the Humboldt fellowship at Heidelberg University is sincerely acknowledged. We acknowledge several fruitful discussions over the years with Drs. J N Goswami, Marc Chaussidon, and Justin Simon. We thank Drs. Mario Trieloff and J.B. Patel for comments that helped bring greater clarity and coherency. Vikram, Sameer, and Alejandro Cisneros helped in preparation of the figure 2 and are thankfully acknowledged. We dedicate this paper to (Late) Prof. Devendra Lal.

Author Contributions

RKM and KKM planned the project and contributed equally towards analysis of the data. KKM performed the ion probe analysis and model calculation. RKM and KKM discussed the results and wrote the manuscript.

Reprints and permissions information is available at www.nature.com/reprints.

The authors declare no competing financial interests.

Correspondence and requests for materials should be addressed to

riteshkumarmishra@gmail.com .

Figure captions

Fig. 1 $^7$Li/$^6$Li vs. $^9$Be/$^6$Li Isotope diagram

(a) $^7$Li/$^6$Li ratios measured in melilite including cosmogenic correction are plotted against measured $^9$Be/$^6$Li in a Type B1 CAIs (E40) from Efremovka (CV~3.1-3.4) chondrite. The two sigma errors of individual data and the error envelope of the isochron regression line obtained using Isoplot 4 (model 1)[32] are also shown. Y axis on right hand side shows the ratios as deviations in ‰ (parts per thousand) from the chondritic (12.02) value. The dotted horizontal line shows the chondritic $^7$Li/$^6$Li ratio. The solar wind composition ($^7$Li/$^6$Li =31 ±4) and cosmogenic isotope ($^7$Li/$^6$Li ~2) ratios, will plot outside the frame on the top left and bottow left corners, respectively (See methods for details).

(b) $^{10}$B/$^{11}$B vs. $^9$Be/$^{11}$B Isotope diagram

The $^{10}$B/$^{11}$B ratios measured in melilite are plotted against $^9$Be/$^{11}$B in a type B1 CAI (E40) from Efremovka. The two sigma error bars of individual data and the error envelope of the isochron computed using Isoplot 4 for E40 are also shown. Y axis on right hand side shows the ratios as deviations in permil (‰, parts per thousand) from the chondritic (0.2481) value. The dotted horizontal line shows the chondritic $^{10}$B/$^{11}$B ratio. The dash dot line and solid gray line indicate the $^{10}$Be/$^9$Be ratio in canonical CV CAIs (8×10$^{-4}$) and the maximum $^{10}$Be/$^9$Be ratio (1×10$^{-2}$) observed in an Isheyevo (CH/CB type) CAI, respectively.

Fig. 2

Ratio of modeled to the inferred Solar System Initial (SSI) ratios of the short-lived nuclides $^{41}$Ca (blue), $^{26}$Al (red), $^{53}$Mn (dark blue), $^{10}$Be (green), and $^{36}$Cl (green) relative to their reference isotope and normalised to $^7$Be has been plotted as a function of irradiation duration (years). The spectral index (γ) of the Solar energetic particle (SEP) in the power-law representation, dN α E$^{-γ}$ dE is adjusted to ~2 so as to constrain production of the observed abundance of $^7$Be/$^9$Be and $^{10}$Be/$^9$Be in the two studied CAIs. The targets are assumed to be spherical and solar (≈ Carbonaceous Ivuna meteorite like) in composition, with sizes varying from 10 μm to 1 cm and following a power-law size distribution (dn/dr α r$^{-β}$; β = 4). Irradiation with the enhanced flux for a few years produces observed $^{7,10}$Be along with $^{53}$Mn, $^{41}$Ca while $^{26}$Al and $^{36}$Cl are under-produced by two orders of magnitude. Refer to supplementary material for addition details.

Tables:

Table 1: Li-Be-B isotope data:

Measured isotope ratios and calculated cosmogenic corrected excesses are shown in the table. Errors are 2σ.

Fig. 1

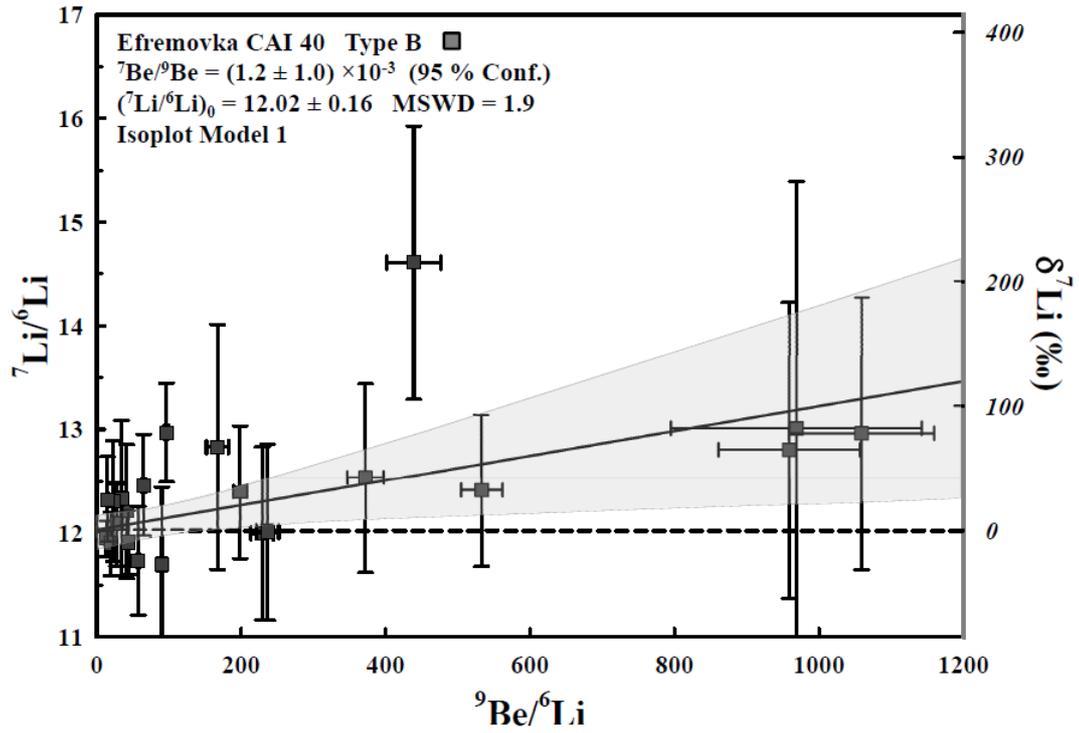

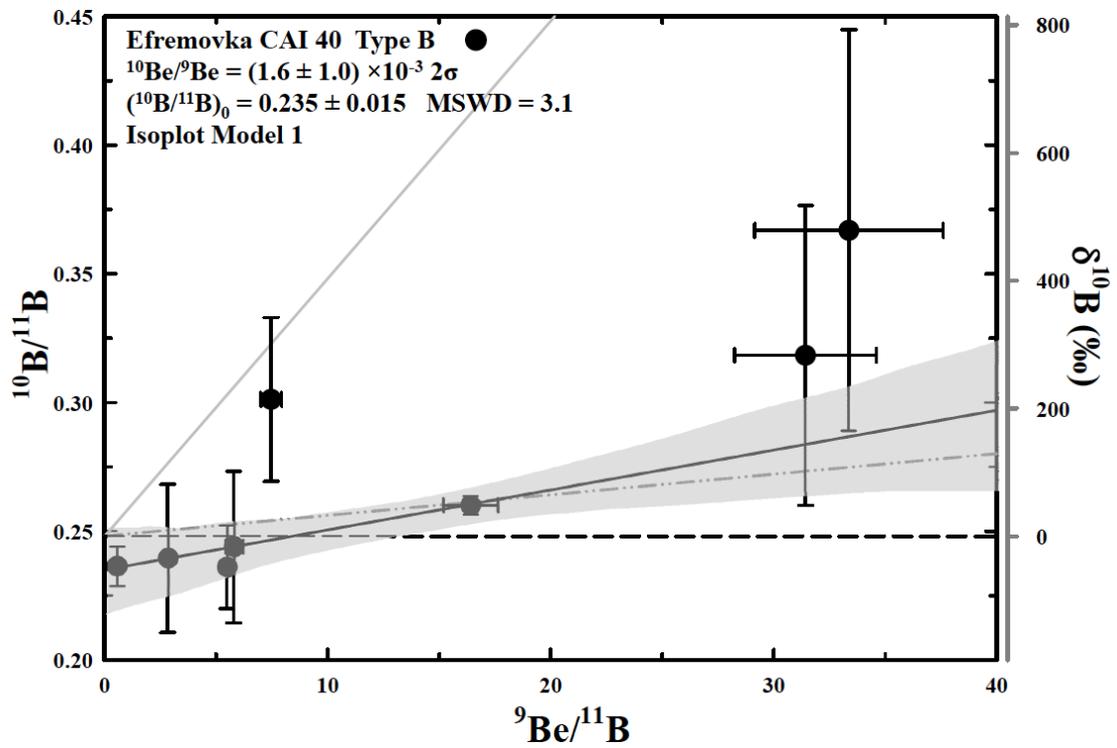

Fig. 2 Plot showing results from model calculations:

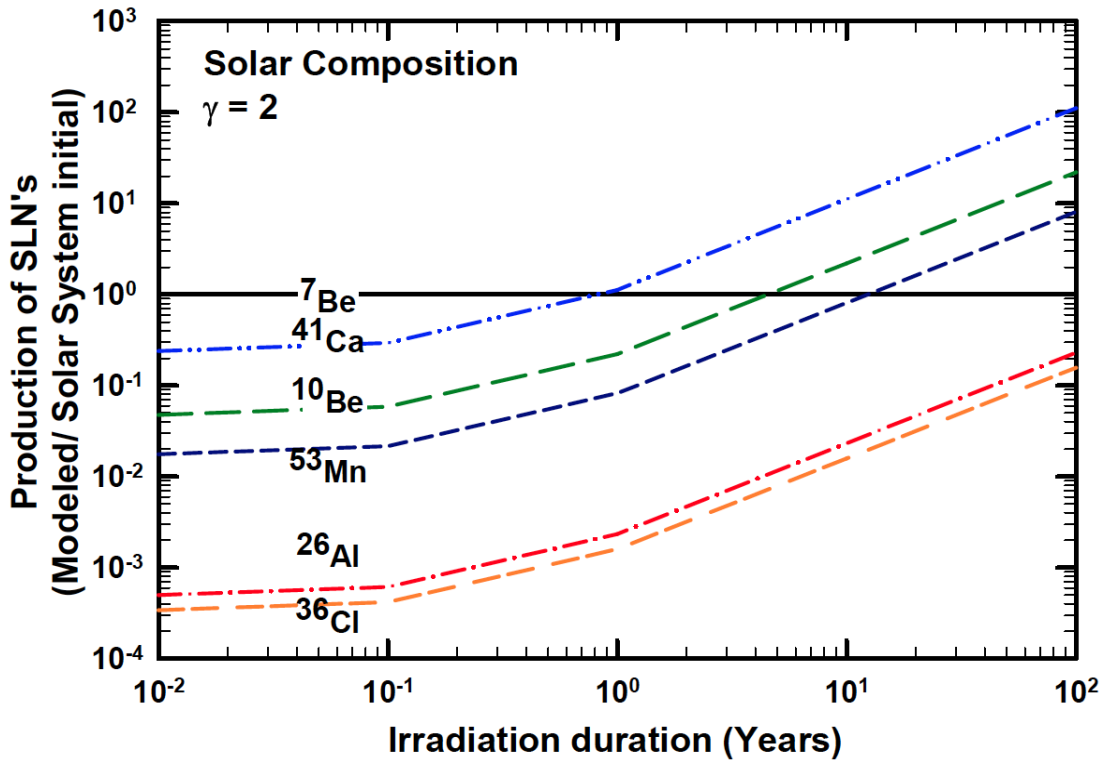

Methods

Sample and Analytical Technique:

Abundance, median size, modal mineralogy, and relative abundance of different petrographic type (A, B, C, FoBs etc.) of CAIs are distinguishing and unique properties of the several distinct groups of meteorites. Carbonaceous chondrites of Vigarano (CV) type hosts CAIs with a high abundance (~8 volume %) and the typical large (~cm) sized Type B CAIs are found in them[16].

Meteorite type and classification:

Efremovka, belonging to reduced CV type, is a 'find' from Efremovka state farm, Palvodar, Russia where a total mass of 21 kg was recovered in 1962. It is petrographically classified as ~3.1-3.4[18].

Efremovka CAI 40: Petrography and Mineralogy:

E40 is a Type B1 broken (less than half) fragment of CAI measuring ~6.4×4.8 mm (Supplementary Figure 1 A-C). Its petrology and mineralogy has been described in detail previously[19] (Supplementary Table 1). It primarily consists of melilite, fassaite (Ti-rich pyroxene), and spinel. Anorthite and pervoskite abundances are rare. Melilite (Gehlenite $Ca_2Al(AlSiO_7)$-$Ca_2Mg(Si_2O_7)$ Akermanite (Åk) solid solution) present in the CAI show a zoned characteristic with Åk content ranging from ~8-72 mol% from outer towards inner region (See Fig. 3 of Ref. 19; Supplementary Figure 1). Typical melilite grains are a few mm in size. Euhedral to subhedral spinel of ~1-500 μm are hosted mostly in the core region with majority of them in the size range of ~50-200 microns. Spinel grains present closer to the periphery, and in the Wark-Lovering (WL) rim are typically smaller in size up to ~50 microns. Spinel (($Mg,Fe)Al_2O_4$) grains are Mg end member in composition (MgO ~ 28.5; $Al_2O_3$ ~70; $TiO_2$ ~0.5-2.0 wt.%; Supplementary Table 1). Pyroxenes mostly, present in the core region, show strongly zoned characteristics in $TiO_2$ abundance and host medium to large sized spinel grains within them. Ca-rich pyroxenes show variation in $TiO_2$ from ~4-14 wt. % (Supplementary Table 1, See also Table 1b of Ref. 19). The composition of melilite, pyroxene, and spinel determined independently during this study are similar in composition to those previously reported (Supplementary Table 1 and Table 1a, 1b of ref. 19). Rare, irregular shaped, anorthites, typical size ~50-100 micron across, are associated with the pyroxenes in the core region. Several Fe,-Ni, metal blebs of up to ~75 micron in size are also present within the central part of the CAI. Modal abundances, size, textures of mineral phases (particularly melilite, pyroxene, spinel), their mineralogy (eg., Åk content of melilite, Ti content of pyroxene, Mg end member spinels), mineral phases associations (eg., anorthite with pyroxene) provides key information about formation of the CAI and also any later stages alteration events seen by it. These petrography and mineralogical features allow to constrain that the E40 CAI crystallized from a melt with a peak temperature of ~1420 ºC (±10) and cooled slowly at ~0.5ºC/hr.

Abundances of Beryllium, Lithium, and Boron: Pattern, crystallization, and comparison:

(a) Beryllium: Beckett et al. 1990 experimentally investigated partition of Be between crystal and melt in melilite[33,34]. Be is a moderately incompatible in gehlenitic melilite with Dmelilite/melt =0.5 for Åk content of 30 % but is compatible in akermanitic melilite of 75 mol % with Dmelilite/melt =1.9[33,34]. Under closed-system crystal fractionation of melilite, the partitioning of Be in melilite for a starting initial concentration of Be (20-50ppb) as a function of increasing Åk content can be calculated and is shown as the light gray band in Supplementary Figure 2 a (also see Fig. 4 of ref. 2). The Be abundance in the analysed melilites in E40 CAI are in general agreement but show scatter around the expected trend (mostly greater compatibility) at higher akermanitic content (Supplementary Figure 2-a). In the analysed melilites in E40 CAI beryllium concentrations range from 39-607 ppb (Table 1). In the Allende CAI 3529-41 studied previously for $^{7}Be$, $^{10}Be$ isotopic records, Chaussidon et al. 2006 observed a range of 25-96 ppb in the magmatic (undisturbed) melilite and up to 7588 ppb within the altered/perturbed melilite. It is interesting to note that all the higher abundances of Be (~200-600ppb) within melilite in E40 CAI are observed when Be becomes more compatible with melilite at the higher akermantic content (~72 mol %) contrary to the observed high abundance at low Åk (~30%) in altered/perturbed melilites in the Allende CAI 3529-41 (Supplementary Figure 2-a). A majority of the observed discrepancy in beryllium concentration at the highest Åk content can be explained if the starting bulk beryllium composition of melt was ~50-80 ppb. Alternatively, partition experiments might constrain the observed dispersion in the distribution of Be at the thermal minima of the gehlenite-akermanite (melilite) solid solution.

(b) Lithium: Lithium concentration in the melilite in E40 CAI range from 1-262 ppb (Table 1). Lithium concentration as a function of Åk content of the melilite and [Be] is shown in Supplementary Figure 2 b-c . Lithium is also an incompatible element with Dmelilite/melt = $0.5^{35}$. A clear trend of Li abundance with Åk content is obfuscated by (1) deviations from the expected trend, and (2) more than two orders of magnitude difference in [Li] for a similar high Åk (~72 %) content. This large variation in lithium concentration of the last solidifying crystal is also seen for Be abundance in E40 CAI. However, it is the lower Li concentrations that are bucking the trend which cannot be explained either by alteration or diffusion. Typically in a later stage alteration/ metamorphism event, addition of Li and or diffusion of Li takes places which is contrary to the observed pattern in E40 CAI. The typical expected trend of later stage processes affecting the abundance of elements (Li, B) in CAI was seen/ noted for non-magmatic melilite[2] in Allende 3529-41 and is shown in Supplementary Figure 2 b. In the magmatic melilites in Allende CAI 3529-41, Chaussidon et al. 2006 noted an increasing trend with increasing Åk (~20-65) for a Li concentration of 9-447ppb. In the perturbed melilites in the Allende CAI the lithium concentration span even wider (order of magnitude higher) range from 1ppb-3.24ppm but with the higher concentration at lower Åk content (~25-40 %). The $^7$Li enriched isotopic composition of the present lithium at high Åk content in E40 CAI cannot be explained either by addition of chondritic lithium or any thermal/ diffusive process. Lithium abundance in melilite does not show any systematic correlation with [Be] (Fig. 2c). It is interesting however to note that despite spanning a similar range of lithium concentration, most of the analysed melilite in the present study have lower [Li] majorly accompanied by higher abundance of Be. In comparison to the E40 CAI, magmatic melilite in Allende 3529-41 CAI display a range of 5-500ppb for a very narrow range of Be concentration up to 0.096ppm. A higher abundance of Be accompanied by lower Li concentration results in higher Be/Li ($^9$Be/$^7$Li) ratio that facilitates detection of in situ decay of $^7$Be.

(c) Boron: Boron, a volatile element, is an incompatible elements in melilite crystallizing from type B CAI composition melt and is evidenced by partition experiments that yield Dmelilite/melt = 0.22[35]. The abundance pattern of B within melilite in E40 CAI shows lack of any systematic dependence on Akermanite content (Supplementary Figure 2 d). The observed abundance of boron increases with [Be] but with significant scatter. The abundance of boron ranges from 4-632 ppb for a similar Åk content of ~ 72%. It is necessary to emphasis here that a small data set of boron in the selected few regions within the melilite in CAI E40 is presented here. A similar range of boron concentration was observed for magmatic melilites in Allende CAI. In summary, the abundances, and pattern/ correlations of beryllium, lithium, boron suggest E40 hosts unaltered, high refractory (relatively) characteristics that makes it a suitable/appropriate sample to study Li-Be-B isotopic records.

Mg isotopic Fractionation and implications:

Mg isotopic composition of melilite show a decreasing fractionation trend (F(Mg) = $\delta^{25}$Mg in ‰/amu) of ~5 ‰ over a distance of ~1000 microns (See Fig. 3,5 of Ref. 19) before attaining a nearly uniform value of ~0 ‰. The decreasing Mg fractionation trend accompanied by decreasing akermanitic composition can be best explained by volatilization of Mg (and other elements) in the outer margin during a brief high temperature event[19]. The 50% condensation temperature of Mg is 1336 K compared to 1653, 1517, 1452, 1310, 1142, 908 K of Al, Ca, Be, Si, Li, and B, respectively[36]. Thus, evaporation of Mg should be accompanied by near complete loss of boron and lithium and proportional portion of silicon in the outer regions. The abundance of refractory Ca, Al, and Be would remain mostly unaffected. The presence of WL-rim, discussed in the following section, at the outer margin also lends credence to the evaporative loss in the outer margins scenario.

Wark-Lovering rim: Characterization and implication for formation and preservation of initial character:

A Wark-Lovering (WL)-rim of ~ 200 μm width is present along the unfragmented half perimeter of the exposed CAI (Supplementary Figure 1). The Wark-Lovering rim sequence consisting of sequential layers, outward to innermost, of olivine, pyroxene, melilite, and spinel-hibonite layer surrounds the convulated boundary of the inclusion (Supplementary Figure 1). Pyroxene layer displays the characteristic gradation from Ti-Al rich to Al-rich pyroxene moving away from the periphery. Perovskites are relatively abundant and scattered randomly throughout the inner most hibonite-spinel-perovskite layer. Anhedral to subhedral hibonites and spinels show symplectic growth in the inner most layer of WL-rim. Petrographic and mineralogical studies of WL-rims in different petrographic class meteorites suggest that the minerals (eg. melilite, anorthite) in WL-rims are quite amenable to alteration and therefore sensitive indicator of post formation aqueous and thermal metamorphism[37]. Presence of these unaltered minerallic layers in WL-rim, Fe,-Ni metal blebs, mineral composition of phases present across the CAI (E40) provide a strong evidence of absence of any significant alteration in the CAI and corroborate other evidences of the pristine character of the CAI. The petrogenesis, unique physico-chemical features, and sets of conditions that favoured the preservation of the pristine characteristics of the CAI may have proved fortuitous and crucial/critical for the detection of the very short-lived $^7$Be isotopic records.

Rare Earth Element Pattern and bulk oxygen isotopic composition of E40 CAI:

The bulk REE composition of E40 is "Type I kind" with enrichment of ~10× CI in LREE going to ~20× CI in HREE[38]. Type I REE pattern is characterized by rather flat distribution pattern of REE with positive europium (Eu) and negative ytterbium (Yb) anomaly. Such an abundance pattern is interpreted as formation of the object from an unfractionated solar abundance reservoir. The positive Eu anomaly coupled with negative anomaly of Yb implies formation in a more oxidizing condition.

Bulk oxygen isotopic composition of E40 CAI with $\Delta^{17}O$ = -2.66 ‰ is consistent with the REE abundance pattern[39]. An in situ oxygen isotopic composition determination along a traverse and also in mineral phases present within and in the WL-rim could provide interesting corroborative information about the formation conditions and later stage (if any) isotopic exchange between the prevalent environment. However, currently only bulk oxygen isotopic composition data is available.

A cosmogenic exposure age of 11.4±1.7 Ma has been inferred from studies of noble gases in the meteorite[40] and is a key parameter in determining the initial $^7$Li/$^6$Li ratio prior to the moderate duration of exposure seen by the Efremovka parent body.

Analytical Technique:

Li-Be-B isotopic analyses: Li-Be-B isotope systematics was performed using peak jumping, pulse counting mode using electron multiplier of the Cameca ims 4f ion microprobe at Physical Research laboratory, Ahmedabad India. The analytical procedure used for Be-B isotope systematics[4] was modified to include additional measurement of two lithium isotopes.

An O⁻ primary beam of ~15-20nA, accelerated at ~12.5 kV, was focused to a spot size of ~35×25 μm on a sample kept at 4.5kV for a total impact energy of ~17kV. The positive secondary ions of lithium, beryllium, and boron generated by bombardment of the O⁻ beam were energy sorted and filtered (~50eV) using an electrostatic analyser. Isotopes were sequentially measured by varying magnetic field intensity. A typical analytical sequence involved measuring ion intensities at mass 5.8, $^6$Li, $^7$Li, $^9$Be, $^{10}$B, $^{11}$B for 10, 50, 10, 5, 30, and 20 seconds, respectively. Multiple analyses up to three were performed at a few spots while count rates and ratios remained stable. The sum of counts obtained over 100-150 cycles were used to calculate ratios. Dynamic background was measured at mass 5.8 (typical value ~0.003-0.008 cps) and the average value was used for correction. In the region with the lowest Li concentration (~1ppb), typical count rates of ~0.1-1.5 cps were measured for $^6$Li and counted for suitable duration to always yield total counts of >1200. Higher counts of greater than a few cps were measured for all other analyses. Typical background counts in the range of 0.003-0.008 cps are negligible compared to the measured cps of lithium and boron isotopes. The flat polished surfaces of epoxy mounted CAIs were coated with ~20 nm of gold and sputtered for ~15-20 minutes to remove the gold coat and surface contamination before the individual analyses. Peak centering of the $^9$Be mass flat top peak was used to monitor the stability of the magnetic fields and counting at the flat top peak centres of the masses was maintained by monitoring and adjusting periodically (every ~20 cycles) any shift in $^9$Be peak centre. A mass resolving power of ~2000 (M/ΔM) was used that is sufficient to resolve all of the major interfering species (most stringent $^{10}$B and $^{9}$BeH MRP required 1416) including $^{9}$Be from $^{27}$Al$^{+3}$(required MRP 491).

GB4 synthetic glass (SiO$_2$ = 72.94 Al$_2$O$_3$= 15.57, Na$_2$O = 4.56 K$_2$O =4.14, CaO= 0.57, Fe$_2$O$_3$ = 0.52, FeO= 0.27, Total =99.78, [Li] = 384 ppm δ$^7$Li$_0$ of -4.3±0.5 ‰, [B] = 970 ppm δ$^{11}$B of -12.8±0.5 ‰ and [Be] = 11.3 ppm) was used as a standard[2] to monitor the mass fractionation and to calculate yields during different sessions. The relative ion yields (λ = [B/Be]$_{true}$/[B/Be]$_{measured}$ ; and similarly [Li/Be]$_{true}$/[Li/Be]$_{measured}$ ) for $^{11}$B/$^9$Be and $^6$Li/$^9$Be ranged from 2.62-2.95 and 1.25-1.75, respectively, during different sessions. Here, measured refer to the ratio measured using ion probe while true refers to the bulk quantitative analyses using atomic absorption spectroscopy. Mass fractionation for Li and B isotope varied between the different sessions but remained within a narrow range during several days of any session in agreement with previous studies. The mass fractionation α$_{Li}$ (α$_{inst}$ = ($^6$Li/$^7$Li)$_{ionprobe}$/($^6$Li/$^7$Li)$_{absolute}$) and α$_B$ were ~1.01 and ~0.94, respectively. The range of variation in approximately one year of analyses was around 12 ‰ & 5 ‰ with an average around 1.009 and 0.940 for Li, and B, respectively whereas for a particular session this was within 2 ‰. Similar instrumental mass fractionation values for lithium (1.01-1.06) and boron (0.94-0.98) were observed by Chausssidon et al. 2006. A detail study to understand variation of instrumental mass fractionation with akermanite (Åk) content in synthetic glass and natural sample is lacking and would be useful. However, Chaussidon et al. 2006 studied GB4 with several other terrestrial standards (BHVO, UTR2, NBS 612) to note that "matrix effects on the instrumental mass fractionation are not a serious complication for the silicate minerals." And used GB4 glass to correct for instrumental mass fractionation for analyses made on melilite, fassaite, and anorthite. So the difference in fractionation factor with Åk content is expected to be negligibly small compared to the other dominant errors of measurement to make any significant effect on the result and inferences drawn in the paper. However the difference in composition of GB4 and melilite can have systematic difference up to ~ 25% in determination of relative ion yield affecting the calculated Be/Li and Be/B. This difference would however not effect our conclusions of isochron value beyond the quoted errors.

Cosmogenic correction:

The experimentally measured $^7$Li/$^6$Li ratios for each analytical spot were corrected for cosmogenic exposure. An independent calculations, following procedures similar to those previously used[2], were made and verified by calculating the corrections for data obtained by Chaussidon et al. 2006. Cosmogenic correction procedure involves calculating the number atoms of lithium isotopes ($^{6,7}$Li) produced per unit mass by exposure to irradiation (GCR) and subtracting it from the measured lithium isotopes to obtain the cosmogenic corrected lithium isotope ratio. Efremovka has exposure age of 11.4 ± 1.7 Ma and a parent body size of radius ≤20 cm has been calculated[40]. Li concentration at each analytical spot was calculated by calibration to the GB4 standard. The primary current normalized count rate measured on GB4 standard (counts/nA/ppm) is ratioed to the counts seen in the meteoritic sample to obtain the abundance of lithium in the melilite in the CAIs. The counting statistics errors associated with measurement of both standard and sample are added in quadrature to give errors of concentration of the sample (melilite). Cosmogenic corrections to $^7$Li/$^6$Li depend on the concentration of lithium and are quite significant (~60, 95 ‰) for analysis having low (~2,1 ppb) lithium concentration (Supplementary Figure 5). A correction protocol used by Leya, 2011 considering the composition of each spot and also parent body size found differences of only a few (<1) ‰ in all the analysed melilites (spots), except for one perturbed spot (out of 41; #III.18(1)$^c$ see appendix 2 of Leya I, 2011) where the difference was 5.8 ‰, with the correction protocol adopted by Chaussidon et al. 2006 and in the present study[15].

Efremovka has cosmic exposure age of (11.4± 1.7) Ma. Correcting the $^7$Li/$^6$Li isotopic ratios of CAI E40 for the mean, maximum, and minimum exposure age of 11.4 Ma, 13.1 Ma, and 9.7 Ma, respectively yield isochrons (model 1 isoplot) with initial $^7$Be/$^9$Be ratio of (1.2±1.0)×10$^{-3}$ (95 % conf.), (1.3±1.0)×10$^{-3}$, and (1.15±1.0)×10$^{-3}$, respectively. The weighted average of these values is (1.2±0.6)×10$^{-3}$. The obtained slopes of the regression ($^7$Be/$^9$Be ratios) for mean, maximum, and minimum exposure age are only marginally different from each other and indistinguishable within associated errors. Therefore, the $^7$Be/$^9$Be ratio with associated error obtained for mean exposure age provides the most appropriate, reliable value, and one that is simultaneously also consistent with the values obtained for the maximum and minimum exposure age. ~15 % uncertainty in cosmic exposure age of Efremovka has however not been added in quadrature to permil level error of individual isotopic measurement.

Diffusion, mixing line, cosmogenic correction, and meaning of correlation:

The isotopic ratio of $^7$Li/$^6$Li after cosmogenic correction displays a linear relationship with $^9$Be/$^6$Li. Apart from the radiogenic decay of $^7$Be, the linear correlation could also result from (1) mixing between two distinct suitable isotopic reservoirs (2) a unique diffusion scenario (3) simply from cosmogenic correction devoid of any relation with Be and/or Li abundance. Or erroneous/ inappropriate cosmogenic correction. Each of these scenario entails an implicit relationship between isotopic ratio and a correlation factor/ parameter and are evaluated in the following discussion. In case of mixing between two reservoirs, one with high lithium concentration and chondritic lithium isotopic ratio and the other with low lithium concentration but high $^7Li/^6Li$ isotopic ratio, a linear positive correlation is expected between the increasing isotopic $^7Li/^6Li$ and decreasing (inverse of) lithium concentration. In the other permutation a negative correlation is expected. In Fig S3 $^7Li/^6Li$ are plotted against 1/[Li] and the following can be observed/ inferred (1) Cosmogenic corrected $^7Li/^6Li$ ratios are rather independent of 1/[Li] but importantly the high lithium isotopic ratio are also displayed by high [Li]. It is important to note though that mean value of at least six spots are higher than chondritic value by more than 64.5 ‰. (2) Measured (uncorrected) lithium isotopic ratios display negative correlation with 1/[Li] with only two spots with high [Li] resolved from the chondritic ratio (Fig S3 b,c). The lithium isotopic ratios of analysis with the lowest [Li] concentration are chondritic within ±15 ‰ (Supplementary Figure 3 b). On the other hand a marginal trend of increasing lithium isotopic ratio both for uncorrected and cosmogenically corrected ratio with [Be] can be seen (Supplementary Figure 4). It should however be noted that significantly large error are associated with several measurements (data points) that makes the correlation statistically poor when considering the errors. The cosmogenic corrections in lithium isotopic ratios in all cases increase the mean isotopic ($^7Li/^6Li$) ratio but by variable amount as noted previously. In summary, there is (1) lack of correlation between $^7Li/^6Li$ with 1/[Li] (2) some correlation of increasing $^7Li/^6Li$ with [Be]. A statistically significant resolved isochron is only obtained between cosmogenically corrected $^7Li/^6Li$ and $^9Be/^6Li$. This kind of correlation is expected in the in situ decay of $^7Be$ scenario that is controlled by Be/Li ratio and not necessarily independently by either [Be] or [Li].

(2) Diffusion Scenario: A mixing between two components can also be induced by diffusion process due to any thermal event. The undetermined, expected fast diffusion rate of lithium makes it favourable for generating the diffusion determined correlation. However, in such a case the previously discussed properties of mixing can be seen. Those criterion and correlation are not observed as has been discussed previously in the mixing scenario case. In addition, orders of magnitude difference in abundance of Li in close neighbourhood that is randomly oriented within the CAI invalidates diffusion as the prime causative factor for the observed correlation. Concurrently though the process is equally efficient in obliterating/ disturbing any previous trends and could have contributed majorly to the observed scatter of the data along the isochron. Given the high temperature formation history of CAIs, and diffusion rates of lithium, only a very rare fortuitous case is expected to show evidence of any kind of isochronal $^7$Be decay.

(3) Cosmogenic corrections as the causative for the observed correlation: Cosmogenic corrections in $^7$Li/$^6$Li are a linear function of lithium abundance (Supplementary Figure 5). The magnitude of corrections are independent of beryllium concentration and also do not show any correlation with the akermanite content. Clearly resolved excesses (>3σ) from the chondritic ratio are present for at least 2 analysed spots without cosmogenic correction. Both these spots have high lithium concentration and cosmogenic correction are insignificantly small of only 10, 13.7 ‰ corresponding to the uncorrected value of 205.4, and 65.4 ‰, respectively. These two spots demonstrate that cosmogenic corrections are not the source of excesses in observed ratio of $^7$Li/$^6$Li. The insignificant cosmogenic corrections in these two spots and a statistically significant correlation demonstrated by the cosmogenically corrected data provides the expected rationale for a significant effect of cosmogenic exposure on altering the ratios predominantly for regions with low abundances.

The linear correlation between cosmogenic corrected $^7Li/^6Li$ and $^9Be/^7Li$ as a result of *in situ* decay of $^7Be$ is further buttressed by the following reasons:

(1) Absence of a high $^7Li/^6Li$ component (2) High diffusivity of lithium isotope

(1) Absence of high $^7Li/^6Li$ component: Galactic cosmic ray spallation of oxygen and carbon produces a component with low $^7Li/^6Li$ ratio of ~2. Mixing between chondritic component ($^7Li/^6Li$ =12.02) and the GCR produced component ($^7Li/^6Li$ = ~2) will yield a mixing line with negative slope and also negative intercept as seen in previous studies[2,7]. Production of an exotic component with $^7Li/^6Li$ ratio significantly greater than ~2 by cosmic ray irradiation is not possible because of same target (oxygen, carbon) and similar proton and alpha particle reaction cross sections and have not been reported. On the other hand, simultaneous production of $^7Be$ with lithium isotopes by spallation reactions on oxygen and carbon is a natural consequence with the production of $^7Be$ governed by conditions and duration of irradiation in the early solar system. Such a simultaneous production of $^7Be$, decaying subsequently to $^7Li$, in regions with low abundance of lithium can give rise to a component with a high $^7Li/^6Li$ ratio.

The solar wind $^7Li/^6Li$ is ~ 31±4 but its low energy (range of few keV) implies penetration depth of less than one tenth of a micron[41]. Hence, direct implantation of solar wind into the CAIs can not explain the isotopic observations. The presence of a Wark-Lovering-rim (>100 um) around both the CAIs implies any directly implanted $^7Li/^6Li$ would have been completely evaporated and lost during the last melting event of formation of WL-rim. Assuming Solar abundance of Li and Be in the solar wind ($^9Be/^6Li$ = 0.3), such a component will plot outside on the top left hand side of Fig. 1.

Mixing between a solar wind component and a chondritic component cannot produce and explain the observed correlation.

(2) High diffusivity of lithium isotopes: Lithium diffusion in pyroxene[27] and feldspar is two to five orders of magnitude faster than other elements (sodium, potassium, magnesium, oxygen, silicon) (Supplementary Figure 6). There exist no experimental data for diffusion of lithium or boron in melilite but to a first approximation similar orders of magnitude difference in diffusion rates between Li and other elements (magnesium, silicon, oxygen) is expected. Thus, the large variation in abundance of lithium by an order of magnitude within a distance of ~200 μm precludes both high temperature and long duration of thermal disturbance after formation of these CAIs. The petrography of E40 CAIs constrains peak temperature of ~1420 °C with a cooling rate ~0.5 °C/hr[26] (See main text). Considering this cooling rate, the CAI would have cooled to ~800 °C in ~1240 hrs, or ~52 days. At 800 °C, the lithium will behave as closed system over a distance of ~200 μm (typical size of analyzed melilites) if we assume similarity in trend of diffusivity of lithium in feldspar and melilite compared with other elements (Supplementary Figure 6). The fast diffusion rate therefore in strictest sense provides a lower bound on diffusion time between the high and low concentration component.

Irradiation Calculation:

Production of $^7$Be in the early Solar system, along with $^{10}$Be, $^{26}$Al, $^{41}$Ca, and $^{53}$Mn due to solar energetic particle interaction (SEP) with nebular material were calculated following the approach described in detail elsewhere[20]. Calculations were carried out by expressing the SEP flux in terms of kinetic energy E, [$dN/dE \propto E^{-\gamma}$], where γ defines the spectral shape. A flux normalization of $N_{E>10\text{MeV}}$ =

100 cm$^{-2}$s$^{-1}$ was considered in these calculations. The flux normalization value represents the long-term (million year) averaged flux of SEPs. Reactions induced by both proton and alpha particles were taken into account assuming an alpha particle to proton ratio of 0.1. Composition of targets (solar/ type B CAI like/ refractory), spectrum of radiation (particle flux) (2-5), total fluence, time duration (yr-Myr), ratio of alpha/proton (0.1-1) are some of the free parameters of the modeling calculations and at times more than one suitable combinations of these parameters can yield the required canonical abundances of the chosen set of SLNs. Irradiation of nebular solids of both CI (≈ solar) and CAI Type B composition was considered and the targets were assumed to be spherical, with sizes varying from 10 μm to 1 cm characterized by a power law size distribution of the type $dn/dr \propto r^{-\beta}$, with β value[20] of 4. The relevant cross sections were taken from the literature data[42-44]. Considering composition of solids to be solar abundance[36], which is equivalent to the composition of carbonaceous chondrite of Ivuna type, the model calculations (Fig. 2) produce best matched observed abundances of $^{7,10}$Be/$^9$Be [($^7$Be/$^9$Be)$_i$ = 1.2×10$^{-3}$, ($^{10}$Be/$^9$Be)$_i$ = 2×10$^{-3}$], under the considered scenario and also calculate simultaneous production of other considered SLNs with half life ≤ 3.5 Ma. The Solar system initial abundances of ($^{41}$Ca/$^{40}$Ca)$_i$ = 1.5×10$^{-9}$, ($^{36}$Cl/$^{35}$Cl)$_i$ = 2×10$^{-5}$, ($^{26}$Al/$^{27}$Al)$_i$ = 5.2×10$^{-5}$ and ($^{53}$Mn/$^{55}$Mn)$_i$ = 6.4 ×10$^{-6}$ have been inferred from various studies[1]. The sources of these SLNs are uncertain and debated (stellar vs. irradiation) amongst a few possible candidates[1,4,5,8,45]. These amongst more than a dozen SLNs are easily produced by n/p reactions and hence have been prime candidates of SLNs produced by irradiation. Several lines of evidence however suggest that $^{36}$Cl (and downward revised abundance of $^{41}$Ca !) is most likely to have been produced by irradiation while $^{26}$Al is majorly sourced from a stellar source. The uncertainties of several parameters used in calculations, when taken into consideration, can vary the calculated abundances by an order of magnitude. Therefore, a corresponding match between calculated abundances amongst different SLNs within a factor of ~10 with the inferred canonical abundances is considered to be a reasonable good fit. The calculated abundances depend significantly on the assumed abundances of the targets (i.e. initial composition of gas/solid), spectral index of particle flux, and reaction cross-sections. Critical evaluation of every parameter and values used in the model calculation and its consequences on the concordance of abundances is beyond the scope of this paper and will be discussed separately. A few previous studies that considered abundance of $^{10}Be/^9Be$ as the fiducial value and calculated expected abundances of other SLNs have evaluated the effect of uncertainties associated with various parameters. Irradiation of solids with CAI Type B composition (higher abundance of Ca, Al, and Be compared to Solar) leads to production of SLNs that are expectedly orders (2-5) of magnitudes larger than canonical.

Data availability.

The data that support the plots within this paper and other findings of this study are available from the corresponding author upon reasonable request.

Supplementary Information:

Supplementary Figure 1

(A) Back scattered electron image of Efremovka CAI 40 (E40). The inset of WL-rim is shown in C. The scale bars are shown at the bottom of the images.

(B) False coloured (Red Green Blue; RGB) mosaic X ray elemental map of E40 where abundance of Mg, Ca, and Al are shown by red, green, and blue colour, respectively.

(C) BSE (upper left panel) and false coloured mosaic images of the WL-rim

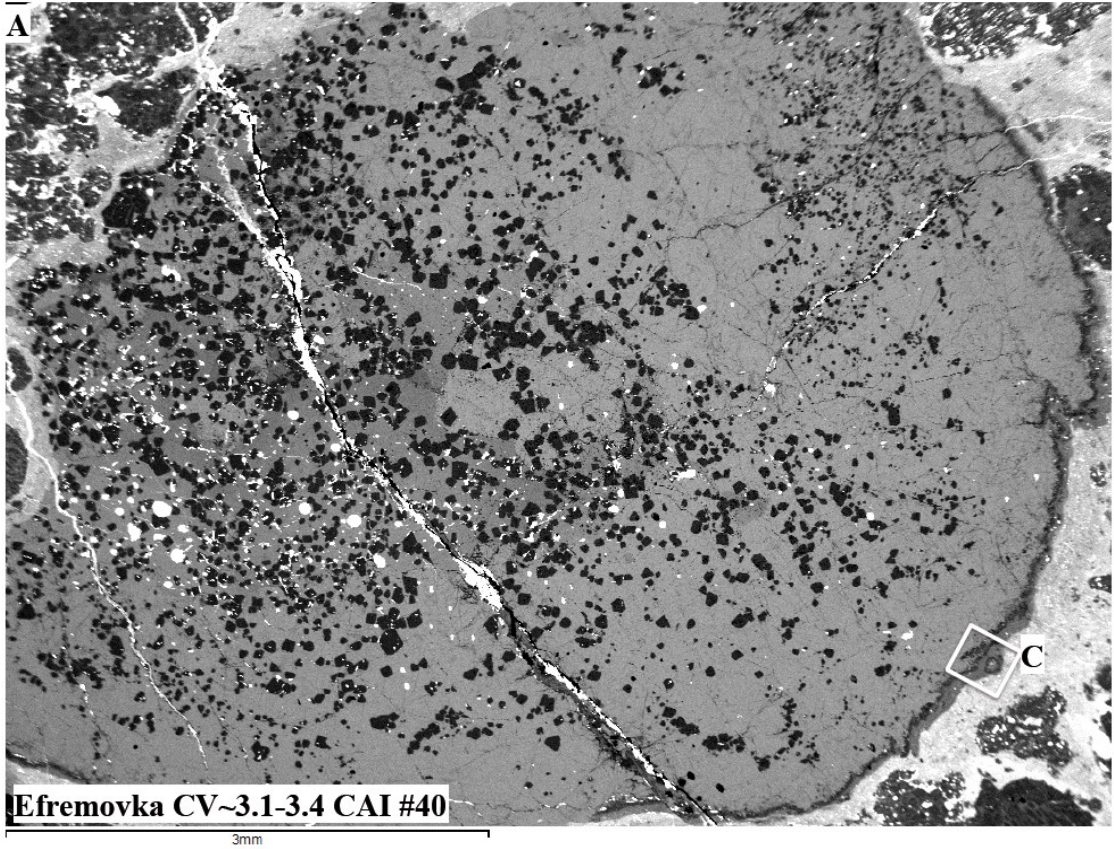
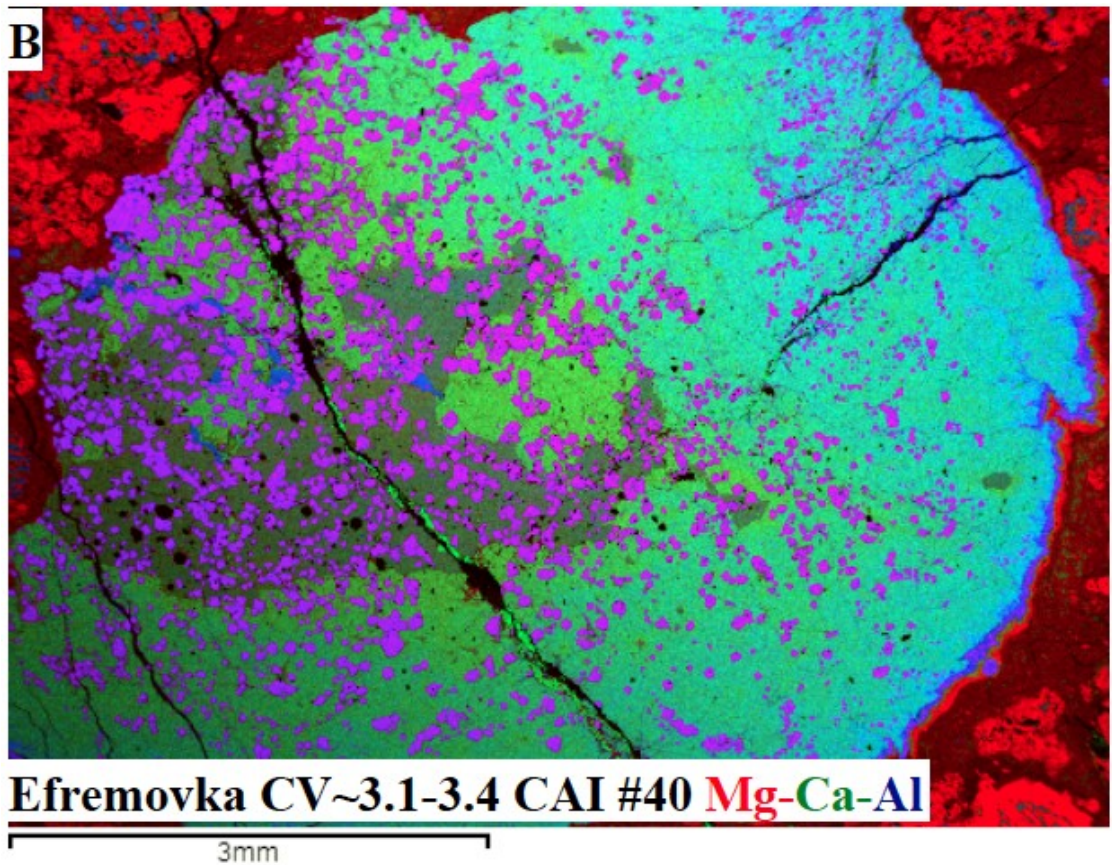

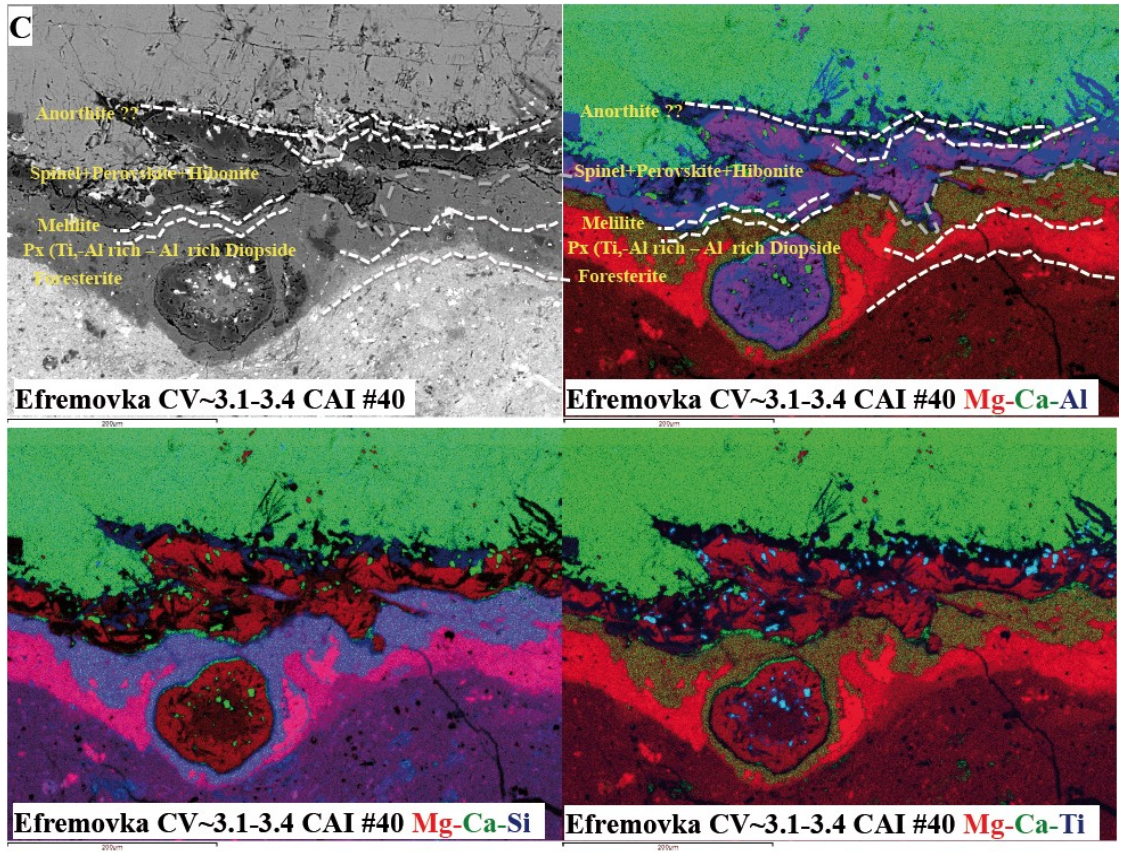

Supplementary Figure 2

(a) Concentration of beryllium as a function of akermanite content in the Efremovka CAI (E40), Allende CAI (3529-41) and Egg-6 CAI are plotted as filled circle, open square, and grey shaded rhombus, respectively. Data for Allende CAI (3529-41) and Egg-6 CAI are taken from Chaussidon et al. 2006 and Spivack et al. 1988. The gray band shows the expected abundance of Be in melilite under a fractional crystallization condition of a type B CAI melt (see supplementary text for detail).

(b) Concentration of lithium as a function of akermanite content in the Efremovka CAI (E40), Allende CAI (3529-41). The filled circle shows data from the present study while open square, and grey shaded square, show lithium abundances in non-magmatic and magmatic melilites in Allende 3529-41 CAI (Chaussidon et al. 2006), respectively.

(c) Concentration of lithium vs. beryllium content in the Efremovka CAI (E40), Allende CAI (3529-41) are shown. The legends are same as in (b).

(d) Concentration of boron as a function of beryllium content in the Efremovka CAI (E40), Allende CAI (3529-41) Legend symbols are same as in (b).

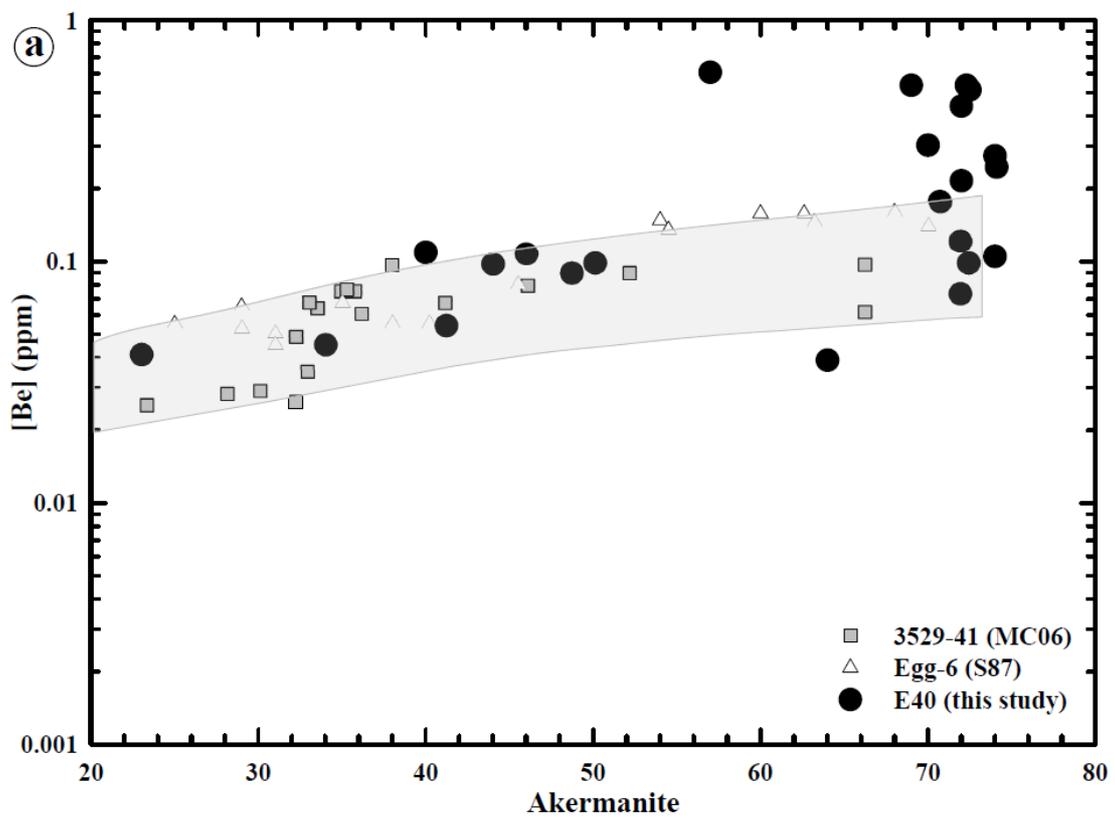

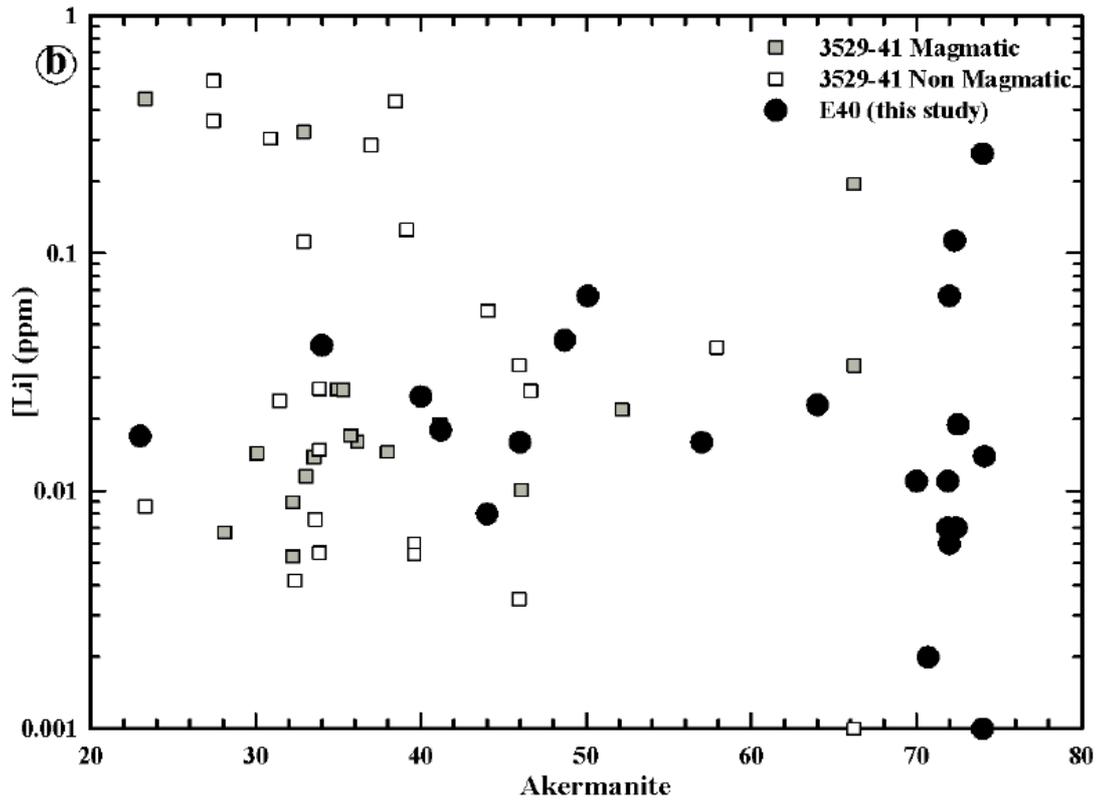

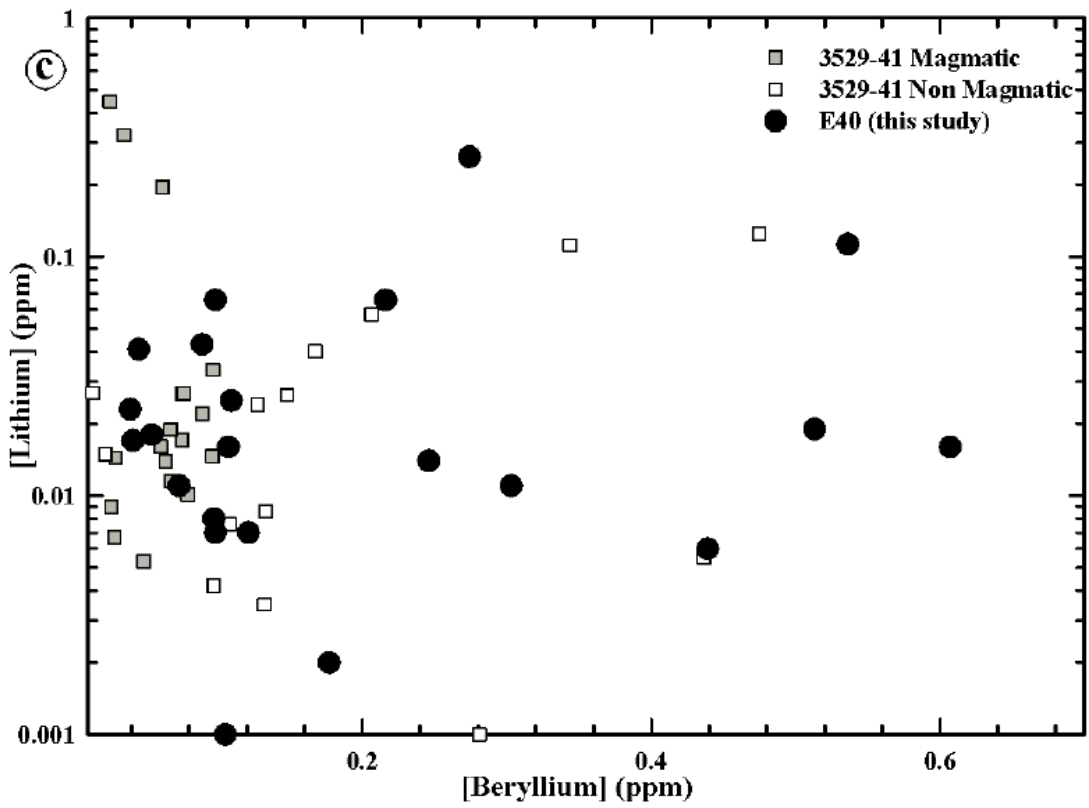

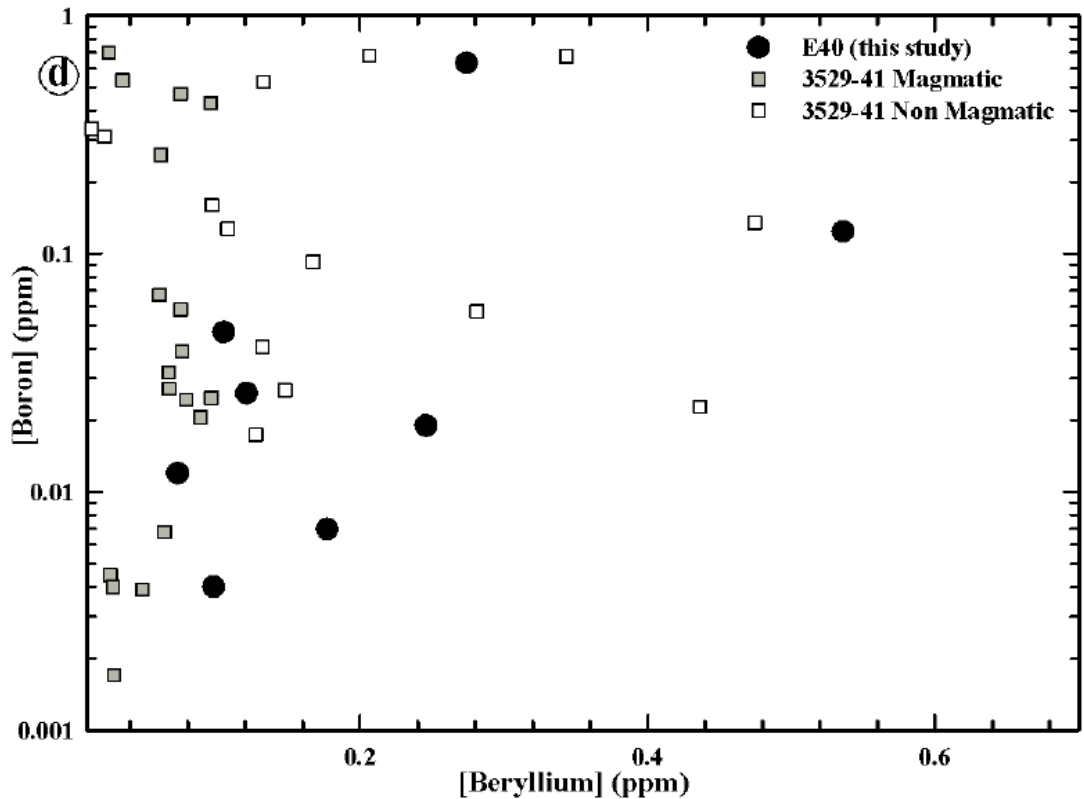

Supplementary Figure 3 Lithium isotopic ratios vs. inverse of lithium concentration diagram

The inverse of lithium isotopic concentration plotted on X-axis with (a) the cosmogenic uncorrected lithium isotopic ($^7Li/^6Li$) ratios, (b) cosmogenic corrected lithium isotopic ratio (c) both cosmogenic uncorrected and corrected $^7Li/^6Li$ ratios (plotted on Y-axis) of data in melilites of Efremovka CAI 40 (E40). The chondritic $^7Li/^6Li$ ratio is indicated by dashed horizontal line. Compare this figure with Fig. 1.

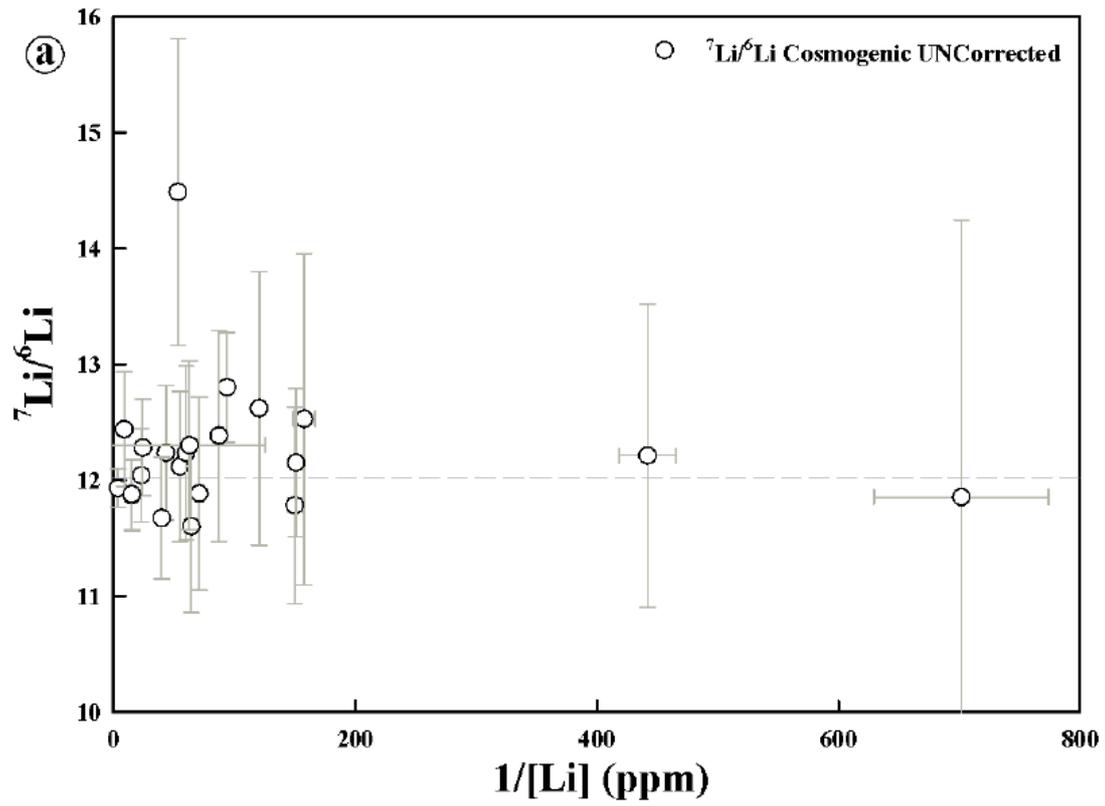
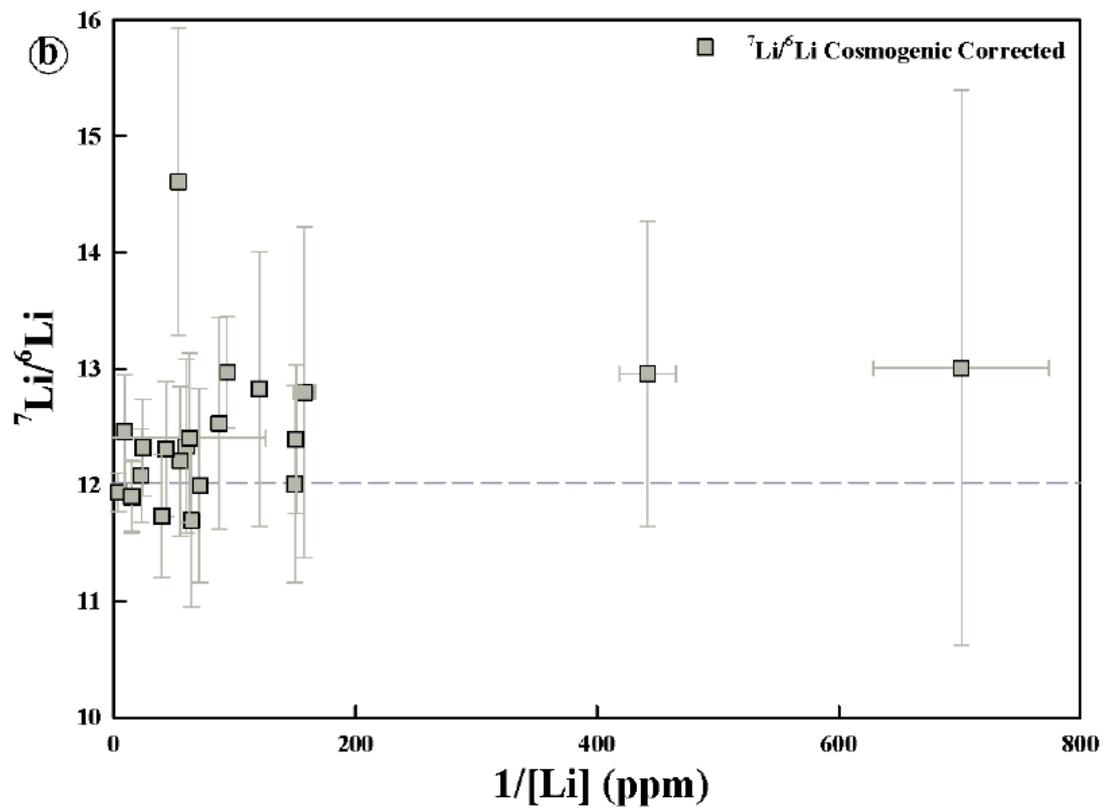

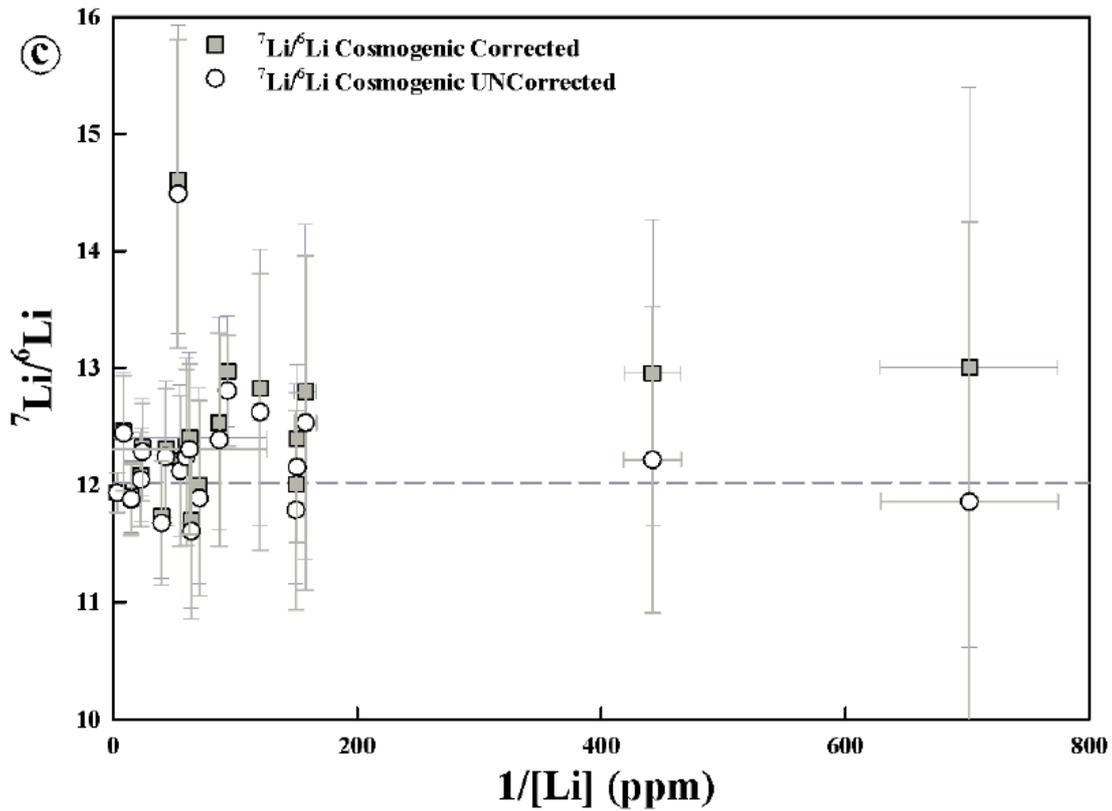

Supplementary Figure 4 Lithium isotopic ratios vs. abundance of beryllium diagram in melilites of Efremovka CAI 40 (E40)

The beryllium abundances in analysed melilite is shown against (a) the cosmogenic uncorrected lithium isotopic ($^7Li/^6Li$) ratios, (b) cosmogenic corrected lithium isotopic ratio. Note Majority of the cosmogenic uncorrected $^7Li/^6Li$ ratio shown in (a) are unresolved from the chondritic. From Fig. (b) it can be seen how the cosmogenic corrections lead to increase (proportional to lithium abundance) in $^7Li/^6Li$ ratio towards yielding a trend of positive correlation. The chondritic $^7Li/^6Li$ ratio is indicated by dashed horizontal line. Compare this figure with Fig. 1.

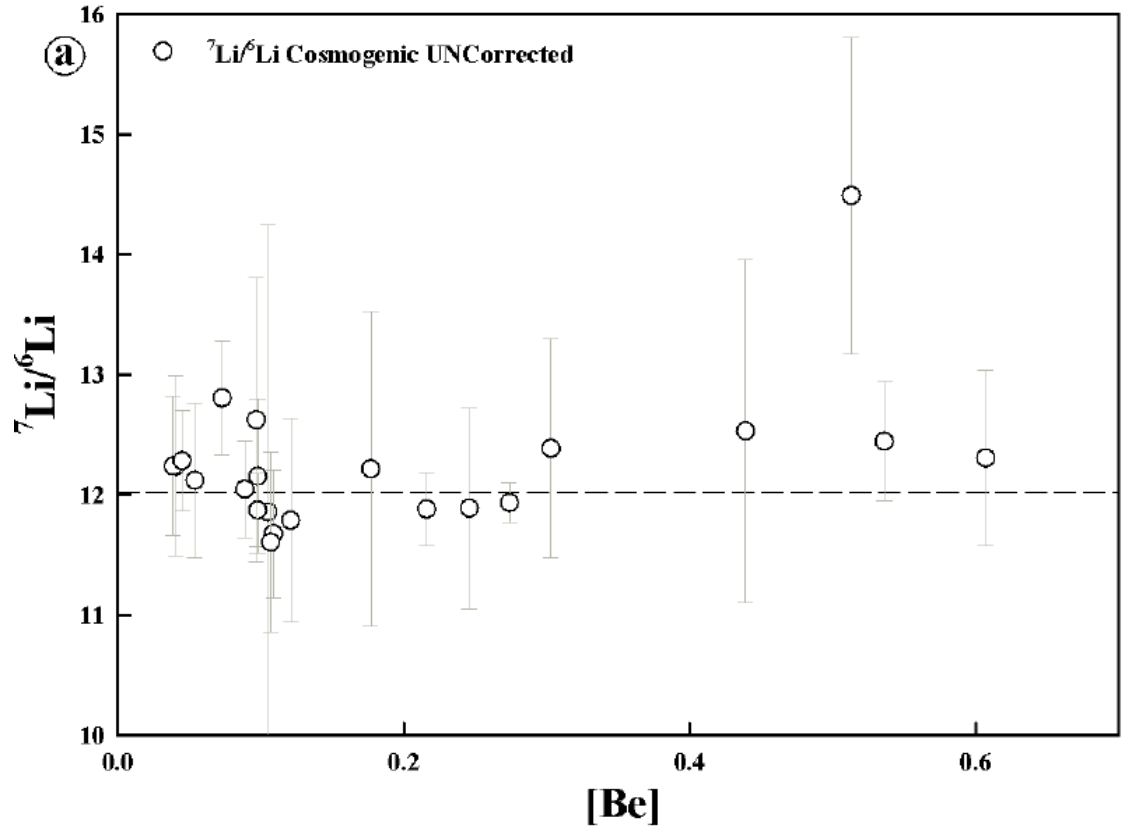
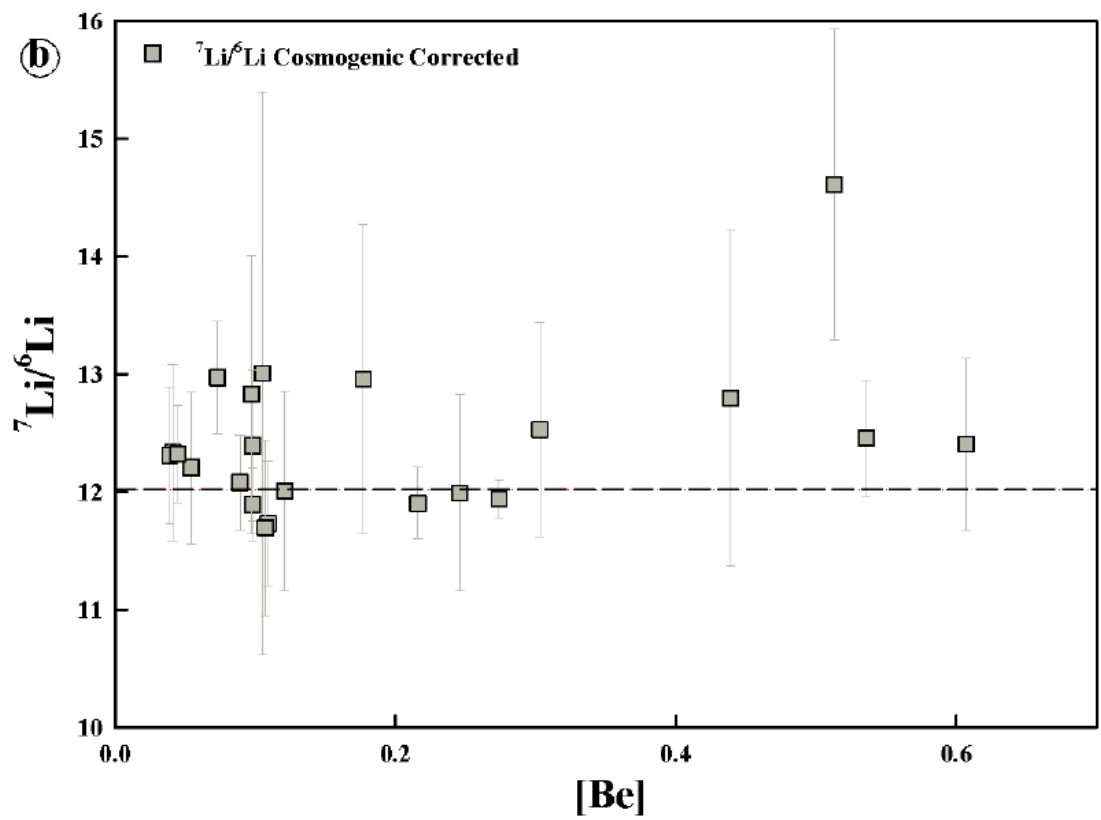

Supplementary Figure 5 Log normal plot of Cosmogenic corrections in $^7Li/^6Li$ ratio vs. abundances of [Li], and [Be] in Efremovka CAI and Allende CAI 3529-41. The closed and open circles shows [Li] and [Be], respectively in E40 while squares unfilled, grey shaded, show abundances of Li in magmatic and non-magmatic melilite. Unfilled half shaded, grey half shaded squares show abundance of [Be] in magmatic and non-magmatic melilite. Note expected correlation of cosmogenic correction with lithium abundance lie on two different lines for Allende CAI (3529-41) and Efremovka CAI (E40) primarily because of different exposure age of 5.2, and 11.4 Myrs, respectively.

(b) Cosmogenic corrections in $^7Li/^6Li$ ratio vs. akermanitic content of the analysed melilite in Efremovka CAI E40. The cosmogenic corrections are nearly independent of akermanitic content. Note a large range of (0-90 permil) cosmogenic correction for high akermanite content (~72 %).

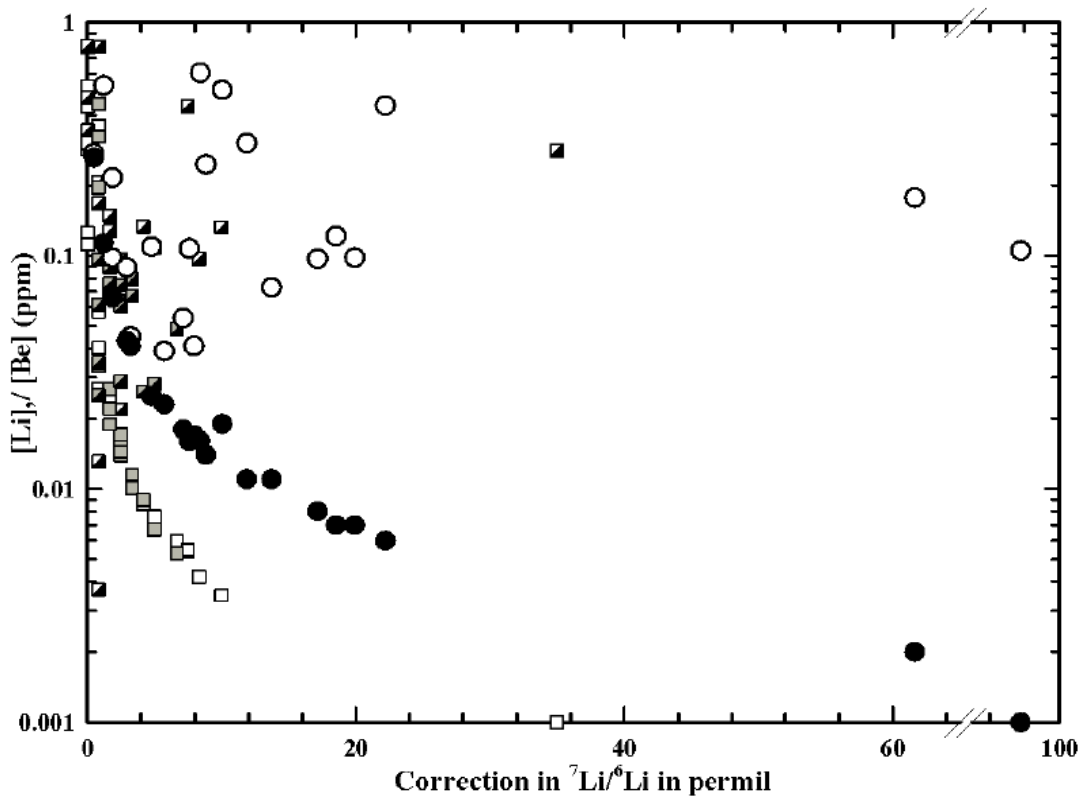

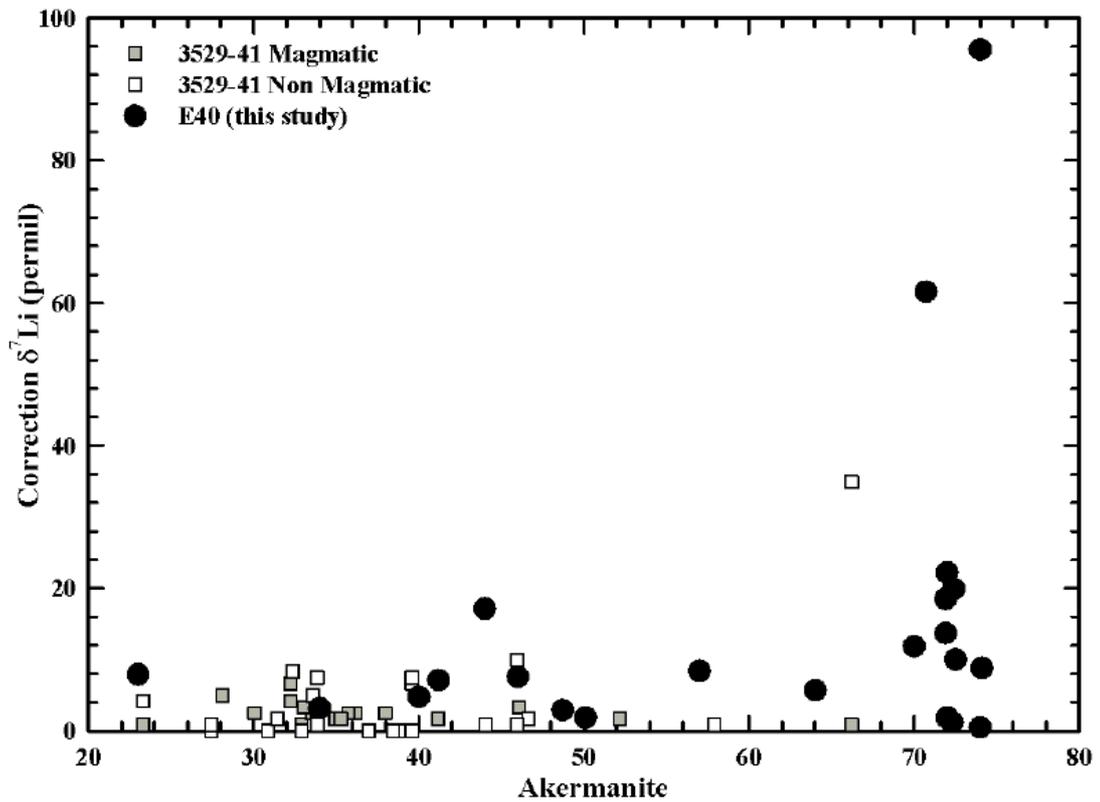

Supplementary Figure 6 Comparison of diffusion time scales for different species (e.g. Ca, Na, K, Mg) in feldspar and melilite for two chosen temperatures of 1400, and 800 °C and diffusion lengths of 50 and 200 μm. The log Do and $E_a$ (activation energy) values are taken from relevant literature[46].

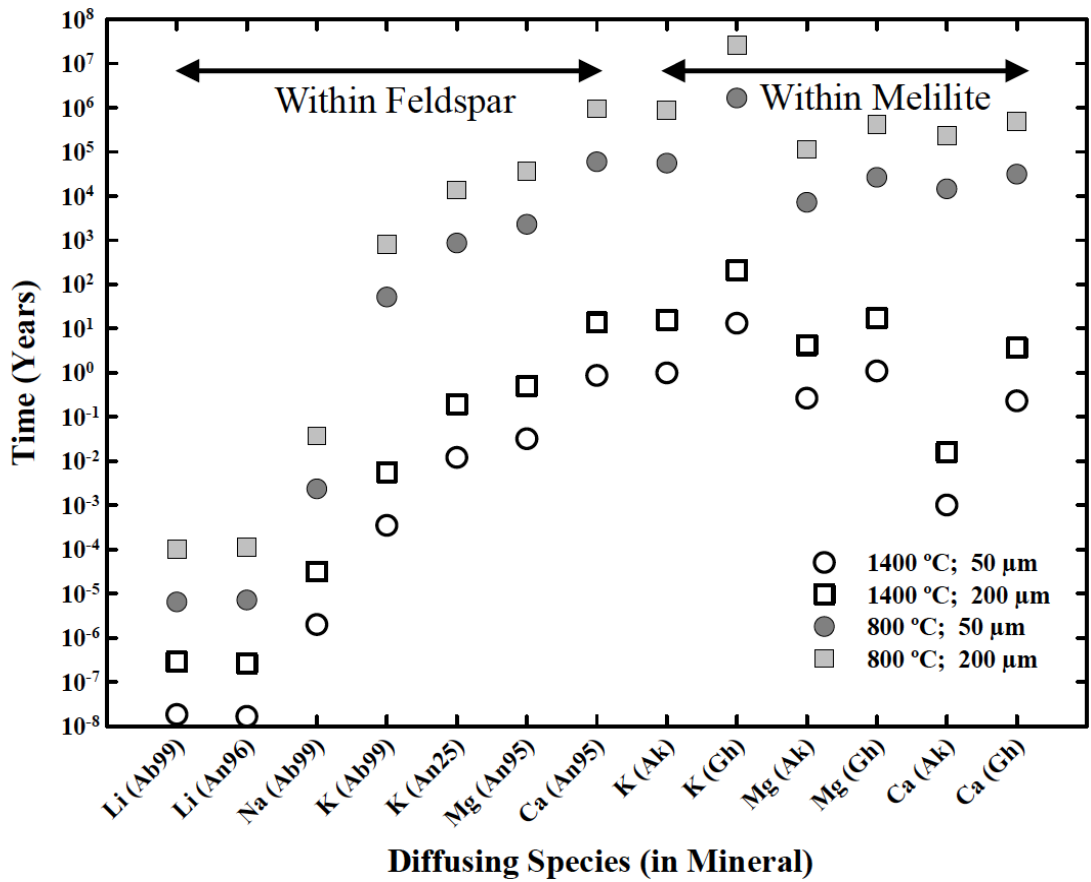

Table 1 Li-Be-B isotopes data in Efremovka CAIs:

| Analysis | $^9Be/^6Li$ | 2σ | $^7Li/^6Li$ GCR Corr. | 2σ | [Li] | 1/[Li] | 2σ | $^7Li/^6Li$ UnCorr. | δ$^7Li$ UnCorr. | $^9Be/^{11}B$ | 2σ | $^{10}B/^{11}B$ | 2σ | [B] | 1/[B] | 2σ | δ$^{10}B$ | 2σ | [Be] | Ak |
|---|---|---|---|---|---|---|---|---|---|---|---|---|---|---|---|---|---|---|---|---|
| 13Dec-E40-1 | - | - | - | - | - | - | - | - | - | 5.48 | 0.22 | 0.23606 | 0.01610 | 0.124 | 8.1 | 0.3 | -45.0 | 65.2 | 0.536 | 69 |
| 13Dec-E40-3 | 968.8 | 173.7 | 13.0064 | 2.3887 | 0.001 | 701.6 | 72.5 | 11.8576 | -13.4 | 2.84 | 0.22 | 0.23949 | 0.02875 | 0.047 | 21.3 | 1.3 | -31.1 | 116.3 | 0.105 | 74 |
| 13Dec-E40-4 | 13.8 | 0.4 | 11.9392 | 0.1662 | 0.262 | 3.8 | 0.0 | 11.9335 | -7.1 | 0.55 | 0.02 | 0.23636 | 0.00767 | 0.632 | 1.6 | 0.0 | -43.8 | 31.0 | 0.274 | 74 |
| 14Dec-E40-1 | 958.9 | 97.9 | 12.7973 | 1.4259 | 0.006 | 157.8 | 9.0 | 12.5305 | 42.5 | - | - | - | - | - | - | - | - | - | 0.439 | 72 |
| 14Dec-E40-2 | 439.1 | 37.6 | 14.6082 | 1.3185 | 0.019 | 53.5 | 2.4 | 14.4876 | 205.4 | - | - | - | - | - | - | - | - | - | 0.513 | 73 |
| 14Dec-E40-2a | 65.3 | 3.2 | 12.4579 | 0.4931 | 0.113 | 8.9 | 0.2 | 12.4435 | 35.3 | - | - | - | - | - | - | - | - | - | 0.536 | 72 |
| 14Dec-E40-3 | 229.6 | 15.7 | 11.9938 | 0.8339 | 0.014 | 71.0 | 2.7 | 11.8878 | -10.9 | 16.40 | 1.22 | 0.25998 | 0.00362 | 0.019 | 55.2 | 3.8 | 47.9 | 14.6 | 0.246 | 74 |
| 19Dec-E40-2 | 1059.0 | 100.4 | 12.9570 | 1.3107 | 0.002 | 442.0 | 23.5 | 12.2163 | 16.4 | 31.43 | 3.18 | 0.31834 | 0.05825 | 0.007 | 141.7 | 14.2 | 287.9 | 235.7 | 0.177 | 71 |
| 19Dec-E40-3 | 236.8 | 15.4 | 12.0088 | 0.8492 | 0.007 | 150.1 | 5.8 | 11.7863 | -19.4 | 5.80 | 0.39 | 0.24389 | 0.02947 | 0.026 | 38.3 | 2.4 | -13.3 | 119.2 | 0.121 | 72 |
| 19Dec-E40-4 | 96.8 | 3.5 | 12.9703 | 0.4755 | 0.011 | 94.0 | 1.9 | 12.8056 | 65.4 | 7.47 | 0.47 | 0.30125 | 0.03177 | 0.012 | 82.2 | 4.9 | 218.7 | 128.5 | 0.073 | 72 |
| 19Dec-E40-6 | 198.7 | 9.6 | 12.3930 | 0.6405 | 0.007 | 151.1 | 4.3 | 12.1537 | 11.2 | 33.38 | 4.23 | 0.36691 | 0.07790 | 0.004 | 272.8 | 34.2 | 484.4 | 315.2 | 0.098 | 72 |
| 29Jun-E40-3 | 19.8 | 0.7 | 11.8944 | 0.3060 | 0.066 | 15.1 | 0.2 | 11.8722 | -13.9 | - | - | - | - | - | - | - | - | - | 0.098 | 50 |
| 29Jun-E40-4 | 27.9 | 1.2 | 12.0809 | 0.4009 | 0.043 | 23.0 | 0.4 | 12.0458 | 0.5 | - | - | - | - | - | - | - | - | - | 0.089 | 49 |
| 30Jun-E40-1 | 41.4 | 2.6 | 12.2066 | 0.6445 | 0.018 | 55.2 | 1.5 | 12.1209 | 6.8 | - | - | - | - | - | - | - | - | - | 0.054 | 41 |
| 30Jun-E40-2 | 34.6 | 2.5 | 12.3334 | 0.7499 | 0.017 | 60.1 | 1.7 | 12.2382 | 16.5 | - | - | - | - | - | - | - | - | - | 0.041 | 23 |
| 30Jun-E40-3 | 23.2 | 1.5 | 12.3085 | 0.5822 | 0.023 | 43.4 | 1.0 | 12.2399 | 16.7 | - | - | - | - | - | - | - | - | - | 0.039 | 64 |
| 1July-E40-2 | 15.3 | 0.8 | 12.3219 | 0.4151 | 0.041 | 24.3 | 0.4 | 12.2834 | 20.3 | - | - | - | - | - | - | - | - | - | 0.045 | 34 |
| 1July-E40-3 | 57.8 | 2.8 | 11.7328 | 0.5280 | 0.025 | 40.0 | 0.9 | 11.6755 | -30.2 | - | - | - | - | - | - | - | - | - | 0.109 | 40 |
| 2July-E40-2 | 167.5 | 15.1 | 12.8296 | 1.1819 | 0.008 | 120.6 | 5.0 | 12.6236 | 48.5 | - | - | - | - | - | - | - | - | - | 0.097 | 44 |
| 2July-E40-3 | 90.5 | 5.9 | 11.6968 | 0.7471 | 0.016 | 64.3 | 2.1 | 11.6057 | -36.0 | - | - | - | - | - | - | - | - | - | 0.107 | 46 |
| 3July-E40-1 | 533.1 | 28.5 | 12.4054 | 0.7287 | 0.016 | 62.9 | 1.8 | 12.3045 | 22.0 | - | - | - | - | - | - | - | - | - | 0.607 | 57 |
| 3July-E40-3 | 372.4 | 25.2 | 12.5290 | 0.9108 | 0.011 | 87.3 | 3.0 | 12.3864 | 28.8 | - | - | - | - | - | - | - | - | - | 0.303 | 70 |
| 3July-E40-4 | 44.0 | 1.3 | 11.9031 | 0.3047 | 0.066 | 15.1 | 0.2 | 11.8807 | -13.2 | - | - | - | - | - | - | - | - | - | 0.216 | 72 |

All errors are 2σ. Concentration of [Li], [B], and [Be] are in ppm.

Supplementary Table 1: Elemental oxide composition of major mineral phases in Efremovka CAI E40 :

| Na$_2$O | SiO$_2$ | MgO | Al$_2$O$_3$ | K$_2$O | CaO | TiO$_2$ | Cr$_2$O$_3$ | MnO | FeO | Total | Åk | Comment |
|---|---|---|---|---|---|---|---|---|---|---|---|---|
| 0.03 | 34.1 | 7.9 | 17.5 | b.d | 41.7 | 0.03 | b.d | b.d | b.d | 101.3 | 53.6 | Melilite |
| 0.05 | 33.3 | 7.2 | 18.8 | b.d | 41.6 | 0.02 | b.d | b.d | b.d | 101.0 | 48.8 | Melilite |
| 0.02 | 24.3 | 1.3 | 34.4 | 0.03 | 41.6 | 0.09 | b.d | b.d | b.d | 101.8 | 8.9 | Melilite |
| 0.03 | 25.0 | 1.9 | 32.7 | b.d | 41.4 | 0.08 | b.d | 0.01 | b.d | 101.1 | 12.5 | Melilite |
| b.d | 27.9 | 3.8 | 28.0 | b.d | 41.5 | b.d | 0.02 | 0.04 | b.d | 101.3 | 25.6 | Melilite |
| b.d | b.d | 28.7 | 70.1 | 0.02 | b.d | 1.5 | 0.19 | 0.02 | 0.05 | 100.7 | - | Spinel |
| 0.02 | 0.1 | 28.6 | 69.6 | b.d | 0.1 | 1.4 | 0.17 | b.d | 0.03 | 99.9 | - | Spinel |
| 0.01 | 42.8 | 11.8 | 15.4 | b.d | 25.5 | 4.1 | 0.07 | 0.01 | 0.08 | 99.9 | - | Pyroxene |
| 0.00 | 36.3 | 8.7 | 17.5 | b.d | 25.2 | 12.1 | 0.07 | 0.05 | 0.19 | 100.0 | - | Pyroxene |
| 0.16 | 43.6 | 0.2 | 35.9 | 0.01 | 20.0 | b.d | 0.04 | b.d | 0.11 | 100.0 | - | Anorthite |
| 0.06 | 43.6 | 0.4 | 36.0 | 0.03 | 19.9 | 0.1 | b.d | b.d | 0.18 | 100.2 | - | Anorthite |

51